\documentclass[twocolumn,showpacs,prc,reprint,nofootinbib]{revtex4}  
\usepackage{epsfig}

\newcommand{\ee}{$e^{+}e^{-}$}
\newcommand{\mumu}{$\mu^{+}\mu^{-}$}
\newcommand{\mee}{M$_{ee}$}
\newcommand{\NN}{$N$+$N$}

\newcommand{\xcc}{$^{12}$C+$^{12}$C}

\newcommand{\arkcl}{$^{40}$Ar+KCl}
\newcommand{\auau}{$^{197}$Au+$^{197}$Au}

\newcommand{\np}{n+p}
\newcommand{\pp}{p+p}

\newcommand{\xdp}{d+p}

\newcommand{\agev}{$A$~GeV}
\newcommand{\gevcc}{GeV/$c^{2}$}

\newcommand{\gevc}{GeV/$c$}

\newcommand{\gev}{GeV}
\newcommand{\mev}{MeV}
\newcommand{\etal}{$et~al.$}

\begin{document}

\title{Dielectron production in Ar+KCl collisions at 1.76\agev}
\author{\footnotetext{corresponding author:}
G.~Agakishiev$^{6}$, A.~Balanda$^{3}$, D.~Belver$^{16}$, A.~Belyaev$^{6}$,
A.~Blanco$^{2}$, M.~B\"{o}hmer$^{12}$, J.~L.~Boyard$^{14}$, P.~Cabanelas$^{16}$, E.~Castro$^{16}$,
S.~Chernenko$^{6}$
% ...
}


\affiliation{
(HADES collaboration)\\
\mbox{$^{1}$Istituto Nazionale di Fisica Nucleare - Laboratori Nazionali del Sud, 95125~Catania, Italy}\\
\mbox{$^{2}$LIP-Laborat\'{o}rio de Instrumenta\c{c}\~{a}o e F\'{\i}sica
Experimental de Part\'{\i}culas, 3004-516~Coimbra, Portugal}\\
\mbox{$^{3}$Smoluchowski Institute of Physics, Jagiellonian University of Cracow, 30-059~Krak\'{o}w, Poland}\\
\mbox{$^{4}$GSI Helmholtzzentrum f\"{u}r Schwerionenforschung GmbH, 64291~Darmstadt, Germany}\\
\mbox{$^{5}$Institut f\"{u}r Strahlenphysik, Forschungszentrum Dresden-Rossendorf, 01314~Dresden, Germany}\\
\mbox{$^{6}$Joint Institute of Nuclear Research, 141980~Dubna, Russia}\\
\mbox{$^{7}$Institut f\"{u}r Kernphysik, Goethe-Universit\"{a}t, 60438~Frankfurt, Germany}\\
\mbox{$^{8}$II.Physikalisches Institut, Justus Liebig Universit\"{a}t Giessen, 35392~Giessen, Germany}\\
\mbox{$^{9}$Istituto Nazionale di Fisica Nucleare, Sezione di Milano, 20133~Milano, Italy}\\
\mbox{$^{10}$Institute for Nuclear Research, Russian Academy of Science, 117312~Moscow, Russia}\\
\mbox{$^{11}$Excellence Cluster 'Origin and Structure of the Universe', 85748~M\"{u}nchen, Germany}\\
\mbox{$^{12}$Physik Department E12, Technische Universit\"{a}t M\"{u}nchen, 85748~M\"{u}nchen, Germany}\\
\mbox{$^{13}$Department of Physics, University of Cyprus, 1678~Nicosia, Cyprus}\\
\mbox{$^{14}$Institut de Physique Nucl\'{e}aire (UMR 8608), CNRS/IN2P3 - Universit\'{e} Paris Sud,
F-91406~Orsay Cedex, France}\\
\mbox{$^{15}$Nuclear Physics Institute, Academy of Sciences of Czech Republic, 25068~Rez, Czech Republic}\\
\mbox{$^{16}$Departamento de F\'{\i}sica de Part\'{\i}culas, Univ. de Santiago de Compostela,
15706~Santiago de Compostela, Spain}\\
\\
\mbox{$^{a}$now at Lawrence Berkeley National Laboratory, ~Berkeley, USA}\\
\mbox{$^{b}$also at ISEC Coimbra, ~Coimbra, Portugal}\\
\mbox{$^{c}$also at ExtreMe Matter Institute EMMI, 64291~Darmstadt, Germany}\\
\mbox{$^{d}$also at Technische Univesität Darmstadt, ~Darmstadt, Germany}\\
\mbox{$^{e}$also at Technische Universit\"{a}t Dresden, 01062~Dresden, Germany}\\
\mbox{$^{f}$also at Dipartimento di Fisica, Universit\`{a} di Milano, 20133~Milano, Italy}\\
\mbox{$^{g}$now at Dipartimento di Fisica Generale, Universit\`{a} di Torino, 10125~Torino, Italy}\\
}


\date{\today}

\begin{abstract}
We present results on dielectron production in \arkcl\ collisions at 1.76\agev.
For the first time $\omega$ mesons could be reconstructed in a heavy-ion
reaction at a bombarding energy which is well below the production threshold
in free nucleon-nucleon collisions.  The $\omega$ multiplicity
has been extracted and compared to the yields of other particles,
in particular of the $\phi$ meson.  At intermediate \ee\ invariant masses,
we find a strong enhancement of the pair yield over a reference spectrum
from elementary nucleon-nucleon reactions suggesting the onset of
non-trivial effects of the nuclear medium.
Transverse-mass spectra and angular distributions have been reconstructed
in three invariant mass bins.  In the former unexpectedly large slopes
are found for high-mass pairs.  The latter, in particular the
helicity-angle distributions, are largely consistent with expectations
for a pair cocktail dominated at intermediate masses by $\Delta$ Dalitz decays.
\end{abstract}


\maketitle

\section{Introduction}
Lepton pairs emitted from the hot and dense collision zone formed
in heavy-ion reactions are appropriate probes for investigating in-medium
properties of hadrons and, in general, the properties of hadronic bulk matter
under extreme conditions.  In fact, according to theory, there is even
potential to detect new phases of nuclear matter in the laboratory by
isolating their telltale signals, among which are the direct decays of the
short-lived vector mesons into \ee\ or \mumu\ pairs \cite{LeupoldMetagMosel}.
Indeed, the electromagnetic current-current correlator, which enters into
the virtual photon emission rate \cite{GaleKapusta}, depends on the
properties of the strongly interacting medium and its constituents
\cite{CBM_physics_book}.

Dilepton spectra measured by the CERES~\cite{ceres_pb_au_158_7per_centrality}
and NA60~\cite{na60_prl} experiments at CERN-SPS energies (40$A$ - 158\agev)
pointed to a significant in-medium modification of the
$\rho$~meson spectral function, as signaled by a large additional yield
(excess) of lepton pairs in the invariant mass region below the
$\rho$~meson pole mass.  At the much lower beam energies of 1 - 2\agev\
the DLS \cite{dls_prl_porter} collaboration at the Bevalac observed a
similar dielectron excess over the ''hadronic cocktail'', i.e.\ the cocktail
resulting from meson decays in the late (freeze-out) phase of the collision.
However, in contrast to the situation at higher energies, for a long time
this excess could not be satisfactorily explained by theoretical
models and became the so-called ''DLS puzzle''.  The baryon-rich regime probed
at low beam energies obviously requires a different theoretical treatment
than the pion-dominated SPS regime.

The excess of electron pairs observed by DLS in the eta mass region has recently been
re-investigated with the HADES\footnote{High Acceptance DiElectron Spectrometer.}
detector \cite{hades_tech} at SIS18 with carbon beams of 1~and
2\agev\ bombarding carbon targets~\cite{hades_cc1gev,hades_cc2gev}.  These
new data fully confirmed the earlier DLS result and thereby re-challenged
theory to come up with a proper description of pair production at low
energies.  Comparing the excitation function of the excess
pair multiplicity with the known $\pi^0$ and $\eta$
meson production \cite{taps_CC,taps_CaCa,taps_thermal} revealed that
the excess scales with bombarding energy very much like the pion yield does,
but not at all like the eta yield.  This finding already suggested baryonic
resonances -- and mainly the $\Delta(1232)$ -- as possible source of the \ee\
excess yield.

Besides the role played by baryon resonance decays, also a strong virtual
bremsstrahlung contribution to the pair yield from mainly \np\ interactions
had been predicted in various microscopic model
calculations~\cite{gale_kapusta,schaefer,shyam1,kaptari}.
Evidently, a good understanding of the $pp \rightarrow ppe^{+}e^{-}$ and
$np \rightarrow npe^{+}e^{-}$ channels is required for a firm
interpretation of dielectron emission from heavy-ion collisions.
First data on dilepton production in nucleon-nucleon collisions had again
been obtained by DLS~\cite{dls_wilson}, although with limited mass resolution
and, at the lowest beam energies, limited statistics.  Initial attempts
to describe those results were based on the soft photon
approximation~\cite{gale_kapusta}, followed later by more involved calculations
using the one-boson exchange (OBE) approach~\cite{schaefer,shyam1,kaptari}.
OBE models include contributions from a number of diagrams which add up
coherently, leading to subtle, but sizeable
interference effects in the cross sections of both \np\ and \pp\
reactions.  Moreover, extending those results to the description of
heavy-ion reactions, e.g.\ in the framework of transport models,
is a very difficult task which is at present not yet fully mastered.

On the experimental side things moved again when HADES started to study
\pp\ and \xdp\ interactions at $E_{kin}$ = 1.25\agev, i.e.\ just below the free
$\eta$ meson production threshold.  The main goal of these experiments was
to understand the \np\ bremsstrahlung contribution to \ee\ production and
to establish an experimental cocktail of dielectrons from ''free'' hadron
decays at SIS18 energies.  Using a deuterium beam, the ``quasi-free''
$np \rightarrow npe^{+}e^{-}$ reaction was therein tagged by the detection
of a forward-going spectator proton.  A comparison of these data with our former
\xcc\ result showed that pair production in the light C+C system can be
understood as resulting from a mere superposition of free
\NN\ collisions~\cite{hades_nn}.  Moreover, the excess pair yield
in the C+C system, when normalized to the pion multiplicity, turned out
to be largely independent of bombarding energy in the range of 1 to 2\agev,
thus providing us with a useful reference spectrum.  Note also that very
recent OBE calculations \cite{shyam3} come very close in describing
the HADES \pp\ and \np\ data consistently.

The questions now arising are: How does the pair yield evolve with
increasing system size?  Does the influence of the hadronic medium set in
and, if yes, what are its characteristic signals?  To address this complex
we present here results from our measurement of \ee\ production in the
medium-heavy reaction system \arkcl\ at 1.76\agev in which we also have
observed $\omega$ production for the first time at SIS18 bombarding energies.
In section II of this article a brief description of the experiment,
as well as of the analysis procedures, is given.  In section III the
reconstructed \ee\ mass distribution is presented and its relevant
features in terms of excess and vector-meson regions are discussed.
In section IV we show and discuss other pair observables: 
transverse-mass and angular distributions.  Finally, section V 
summarizes and concludes the paper.

\section{The Ar+KCl  experiment}

\begin{figure}[!htb]

  \mbox{\epsfig{width=0.90\linewidth, figure=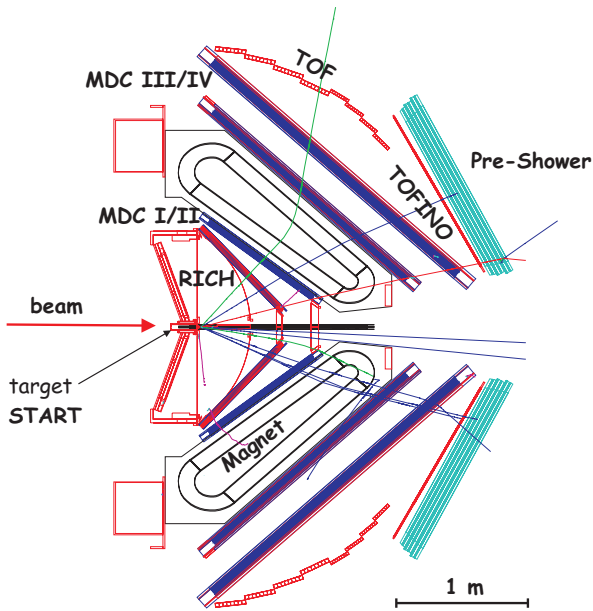}}
  \vspace*{-0.2cm}
  \caption[]{(Color online) Schematic view of the HADES detector as implemented in
    the simulation.  Simulated particle tracks are shown as well.
    }
  \label{hades_cross}
  \vspace*{-0.2cm}
\end{figure}

The HADES detector operates at the GSI Helmholtzzentrum f\"{u}r
Schwerionenforschung in Darmstadt with proton and heavy-ion beams being
provided by the synchrotron SIS18.  Technical aspects of the spectrometer
are described in ~\cite{hades_tech}, a schematic view is shown in
Fig.~\ref{hades_cross}.  Here we recall that its main component serving
for electron and positron identification (PID) is a ring-imaging
Cherenkov detector (RICH).  Further PID power is provided by (1)
the time of flight measured in the plastic scintillator time-of-flight
wall (TOF), (2) the electromagnetic shower characteristics
observed in the Pre-Shower detector, and (3) the energy loss signals from
the four wire-chamber planes and the scintillators of the TOF wall.
A $50~\mu m$ thick segmented polycrystalline diamond detector (START)
placed in front of the target provides the precise event time.

\begin{figure}[!htb]

  \mbox{\epsfig{width=0.90\linewidth, figure=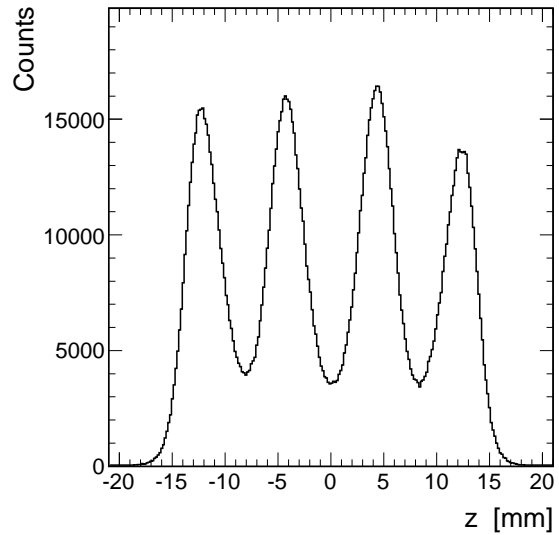}}
  \vspace*{-0.2cm}
  \caption[]{Reconstructed event vertex distribution along the beam axis $z$.
             The four KCl target segments are clearly separated.
          }
  \label{z_vertex}
  \vspace*{-0.2cm}
\end{figure}

A beam of $^{40}$Ar ions with a kinetic energy of 1.756\agev\ was used for
the Ar+KCl experiment discussed in this paper.  The beam intensity was
about $6\times 10^{6}$ particles per 8-second spill.  The four-fold segmented
target was made of natural KCl with a total thickness of 5~mm,
corresponding to 3.3\% nuclear interaction length.  The segmentation of
the target (1.3\% radiation length per segment) helped to minimize the
conversion of photons into electron pairs in the target material.
The four segments are indeed nicely visible in the reconstructed event
vertex distribution along the $z$ axis (i.e.\ beam axis)
shown in Fig.~\ref{z_vertex}.

The on-line event selection was done in two steps:  a first-level
trigger (LVL1) picked out those reactions in which the number of hits
exceeded 16 in the TOF scintillators.  In total, $2.2\times10^9$ of such
events were examined with a second-level trigger (LVL2) to find
single lepton signatures of which about $6\times10^8$ were finally recorded.
In the off-line analysis, events were further filtered by cutting on
the event vertex reconstructed with a resolution of $\sigma_x =
\sigma_y \simeq$ 0.4~mm and $\sigma_z \simeq$ 1.9~mm.
These vertex cuts removed about 5\% of the events in accordance with the
event rate observed in an empty-target run.
The LVL1 trigger enhanced the mean pion multiplicity approximately two times
with respect to the minimum-bias multiplicity.  From our charged-pion
analysis and a simulation of the trigger response to UrQMD events,
we found that this corresponds to a mean number of participating
nucleons of $\langle A_{part}^{LVL1} \rangle = 38.5$ and an average
charged-pion multiplicity of $\frac{1}{2} (N_{\pi^+}+N_{\pi^-}) =
3.50\pm0.12(stat)\pm0.22(sys)$ (for details see \cite{hades_pi,hades_k0s}).

To investigate systematic effects in the dielectron reconstruction of
the HADES experiment, three parallel data analyses were done
with identification algorithms based respectively on (i) a multi-variate
analysis (MVA) \cite{simon_thesis,tmva}, (ii) a Bayesian
approach \cite{martin_thesis,hades_tech}, and (iii) a combination
of multi-dimensional selection cuts \cite{filip_thesis,hades_tech}.
All three analyses led to consistent results with similar signal purities.
The remaining differences were compounded with the other systematic uncertainties
(see below).  Identified electrons and positrons were further combined into pairs.
The total number of reconstructed \ee\ pairs, $N_{+-}$, can be decomposed as
$N_{+-} = S\,+\,CB$, where $S$ denotes the number of signal pairs and
$CB$ stands for the number of combinatorial background pairs.  The former
are the correlated \ee\ pairs originating from the same parent particle
and the latter ones are due to combining uncorrelated leptons stemming from
separate sources, mostly $\pi^0$ Dalitz decays and external photon conversion.
To reduce the $CB$, and hence enhance the $S/CB$ ratio, it is in particular
necessary to suppress contributions from photon conversion, from tracking
fakes, and from misidentified hadrons.  The main source of photons are the
neutral-pion decays, $\pi^0 \rightarrow \gamma\gamma$.  As the induced
conversion pairs have mostly small opening angles, they have been
effectively decimated with an opening-angle cut requiring
$\alpha_{e^+e^-} \ge 9^\circ$ in the laboratory frame.
Tracking fakes were suppressed by selecting only ring-track combinations of
sufficient reconstruction and matching quality \cite{hades_tech}.
Additionally, a single-lepton
momentum cut of $0.1$~\gevc\ $<p_{e}<1.1$~\gevc\ confined the fiducial
acceptance to the region where the combined track reconstruction and lepton
identification efficiency was at least 10\%, but typically 30 - 70~\%,
while the contamination of the lepton sample by charged pions and protons
stayed well below 20\%.

Finally, to obtain the \ee\ invariant mass signal, the remaining CB was
subtracted from all reconstructed pairs in the following way:
in the low-mass region $M_{ee}<0.4$~\gevcc, where the
correlated background from the $\pi^{0}$ two-photon decay followed
by double conversion contributes most, the combinatorial background was
determined using a method based on like-sign $e^{+}e^{+}$ and
$e^{-}e^{-}$ pairs emerging from the same event,
i.e.\ $CB=2\sqrt{N_{e^+e^+}N_{e^-e^-}}$.   For larger masses, however, where
the statistics of like-sign pairs is poor, we used a mixed-event CB
normalized to the like-sign CB \cite{hades_tech}.  The mixing
was done between events belonging to the same event class in terms
of the track multiplicity (five selections) and the target segment
(four selections), i.e.\ for a total of twenty event classes.
This procedure was applied likewise to all other pair observables,
in particular the pair transverse momentum distribution.

\begin{figure}[!htb]

  \mbox{\epsfig{width=0.90\linewidth, figure=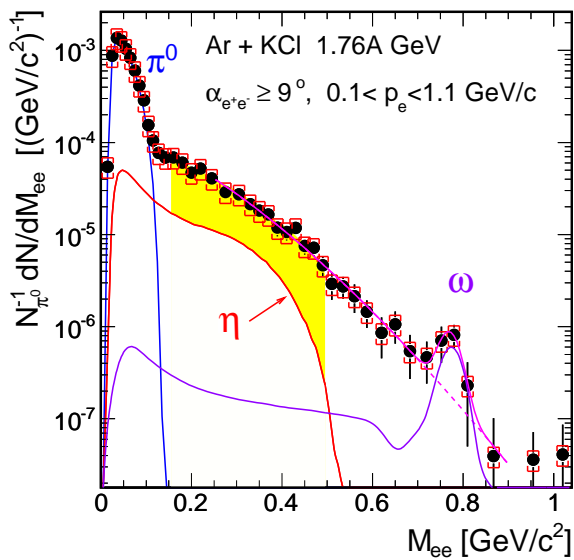}}
  \vspace*{-0.2cm}
  \caption[]{(Color online)
             Reconstructed \ee\ mass distribution in Ar+KCl collisions
             (averaged over three PID analyses, efficiency-corrected, CB subtracted,
             and normalized to $N_{\pi^0}$).  Statistical and systematic errors of the
             measurement are shown as vertical bars and horizontal cups,
             respectively.  Curves represent the $\pi^0$ and $\eta$ Dalitz
             components, as  well as the $\omega$ contribution (Dalitz and direct)
             simulated with the event generator Pluto.
             Also shown are the excess yield over the simulated cocktail
             (shaded area) and a fit (exponential + Gaussian curves) to the
             data in the mass range 0.25 - 0.9~\gevcc\ (see~III.C for details).
          }
  \label{av_mass_spectrum}
  \vspace*{-0.2cm}
\end{figure}

The resulting invariant mass spectrum of the dielectron signal, corrected for the
detector and reconstruction inefficiencies\footnote{Inefficiencies were
  determined with an event overlay technique in which simulated lepton
  tracks were embedded into real events, reconstructed, and tallied.},
but not acceptance, is shown in Fig.~\ref{av_mass_spectrum}.  The spectrum
is normalized to the average number of charged pions -- also measured in
HADES \cite{hades_pi} -- namely $\left(N_{\pi^-}+N_{\pi^+}\right)/2 = 3.5$
per LVL1 event.  As expected from isospin symmetry, this average is a good
estimate of the actual $\pi^{0}$ yield $N_{\pi^{0}}$, i.e.\ we set
$N_{\pi^{0}} = \left(N_{\pi^-}+N_{\pi^+}\right)/2$.
The normalization to $N_{\pi^{0}}$ compensates in first order the bias
caused by the implicit centrality selection of our LVL1 trigger.  The spectrum
shown represents an averaged result from the three parallel PID analyses
mentioned above.  Besides the statistical error bars systematic errors are
represented as horizontal ticks.  They cover systematic effects attributed to
the efficiency correction and combinatorial background subtraction (20\%),
to the error on the normalization (11\%), and to differences between
the three PID methods (10\%).  Statistical errors are of course
point-to-point, the normalization error is global, and the remaining
systematic errors are slowly varying with pair mass.
The systematic errors given are upper bounds and add quadratically to 25\%.


\section{Results from Ar+KCl}

Here we discuss in more detail the efficiency-corrected and CB-subtracted
dielectron invariant mass spectrum from Ar+KCl (see Fig.~\ref{av_mass_spectrum}).
The total yield of $\simeq$85k signal pairs is distributed over three easily
distinguishable regions:  (i) masses below 0.15~\gevcc, dominated by
the $\pi^0$ Dalitz peak, contribute around 70k,  (ii) the intermediate range
of 0.15--0.5~\gevcc\, where the pair excess is located, holds about 15k,
and (iii) masses above 0.5~\gevcc\, where the dileptons from vector meson
decays are expected, total a few hundred pairs only ($\simeq450$).  All pair
observables presented below have been obtained from inclusive LVL2-triggered events,
i.e.\ with no further centrality cuts.  An investigation of different
event classes selected by analysis cuts on the hit multiplicity revealed indeed
a slight dependence of the normalized pair yields on centrality.  However,
as in this still rather small reaction system only a limited range of $A_{part}$
can be scanned via such multiplicity cuts, we discuss below the $A_{part}$
dependence only in the context of a comparison of Ar+KCl with C+C.


\subsection{Low-mass pairs}

\begin{figure}[!ht]

  \mbox{\epsfig{width=0.90\linewidth, figure=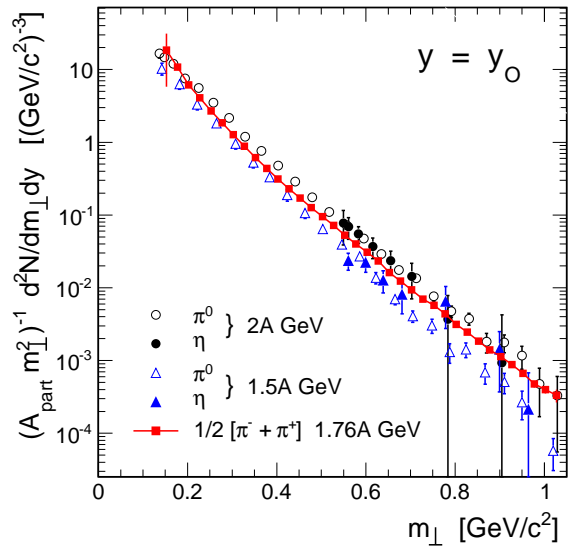}}

  \vspace*{-0.1cm}
  \caption{(Color online)
   Average of the mid-rapidity charged pion $d^2N/dm_{\perp}dy$ distributions,
   $N = 1/2 [N_{\pi^+} + N_{\pi^-}]$, measured with HADES in the 1.76\agev\ Ar+KCl
   reaction \cite{hades_pi,hades_k0s} (squares linked by solid curve), compared
   with the corresponding neutral pion and eta data from
   1.5\agev\ $^{40}$Ar+$^{nat}$Ca and 2.0\agev\ $^{40}$Ca+$^{nat}$Ca
   reactions obtained with the photon calorimeter
   TAPS \cite{taps_CaCa,taps_thermal} (circles and triangles).
   All yields are normalized to their respective $\langle A_{part} \rangle$;
   error bars shown are statistical.
}

   \label{mt_taps}
   \vspace*{-0.2cm}
\end{figure}

The low-mass region contains the bulk of the pair yield, but it is also
the one most strongly affected by the momentum-dependent efficiency corrections.
As more than 90\% of this yield stems from the $\pi^0$, it offers a convenient
handle for validating our dielectron reconstruction and normalization procedures.
To do this check we simulated the pion and eta contributions to the \ee\ cocktail
with the Pluto event generator \cite{pluto,pluto2} using the meson multiplicities
and source parameters given in Tab. \ref{pluto_params}.  In case of
the $\pi^0$ they were taken from our own charged-pion data \cite{hades_pi,hades_k0s},
in case of the $\eta$ they were interpolated from TAPS two-photon measurements
of 1.5 and 2.0\agev\ Ar+Ca (Ca+Ca) reactions \cite{taps_CaCa,taps_thermal}.
Note that for both mesons a two-slope parameterization has been used.
The validity of this interpolation is confirmed in Fig.~\ref{mt_taps} where
$\pi^0$ and $\eta$ mid-rapidity $1/m^2_{\perp} d^2N/dm_{\perp}dy$ spectra from TAPS
are compared with the corresponding $\pi^+$ and $\pi^-$ average measured
in Ar+KCl \cite{hades_pi} (mid-rapidity $y_0 = 0.858$ and $|y-y_0|<0.1$).
In this figure, the different centrality selections of the TAPS
(minimum bias) and HADES (LVL1) experiments are compensated by
normalizing to the mean number of participating nucleons
$\langle A_{part} \rangle$ = 20 and 38.5, respectively.
The Ar+KCl charged-pion average falls nicely in between the neutral pion
(and eta) data, as expected from the smooth beam energy dependence of pion
production.  The characteristics of the vector meson sources ($\rho, \omega, \phi$)
are still largely unknown and have tentatively been set
as given in Tab.~\ref{pluto_params}.

\begin{table}[htb]
  \caption[]{Thermal source parameters used in the Pluto simulation
    of the dielectron cocktail.  For all listed particles we give the
    multiplicity ($N$) within our LVL1 trigger condition, the source
    temperatures ($T_1$ and $T_2$), the relative strength
    ($frac = c_1/(c_1+c_2)$, where $c_1$ and $c_2$ are the amplitudes of
    the two components) of the first component, the polar anisotropy ($A_2$), and the
    helicity coefficient ($B$) of the dielectron decay (see Sec.~III.E).
  }
             
  \vspace*{0.2cm}
  \begin{center}
  \begin{tabular}{ l | c c c c c c }
  \hline
  Part. & ~$N$~ & $T_1$ [\mev] & $T_2$ [\mev] & ~$frac$~ & ~$A_2$~ & ~$B$~\\
  \hline
  \hline
  $\pi^0$ & 3.5 & 52 & 89 & 0.85 & 0.75 & 1\\
  $\eta$ & $8.8 \cdot 10^{-2}$ & 52 & 89 & 0.85 & 0.75 & 1\\
  $\Delta^{+,0}$ & 3 $N(\pi^0)$ & 80 & - & 1 & 0.75 & 1\\
  $\omega$ & $6.7 \cdot 10^{-3}$ & 80 & - & 1 & 0.75 & 0\\
  $\rho$ & $5 \cdot 10^{-2}$ & 80 & - & 1 & 0.75 & 0\\
  $\phi$ & $2.6 \cdot 10^{-4}$ & 80 & - & 1 & 0.75 & 0\\
  \hline
  \end{tabular}
  \end{center}
  \label{pluto_params}
\end{table}

Turning back to Fig.~\ref{av_mass_spectrum}, one can see that the low-mass
pair yield is indeed described very well by our calculation, lending
confidence to the reconstruction process.

\subsection{Intermediate-mass excess}


Comparing a Pluto simulation of long-lived sources (i.e.\ emitting
mostly after freeze-out) with the data in Fig.~\ref{av_mass_spectrum} reveals
that also in Ar+KCl the contributions from $\eta$ (and $\omega$) Dalitz
decays do not exhaust the measured pair yield at intermediate masses,
i.e.\ for $M_{ee} \simeq 0.15-0.5$ \gevcc. 
Just like in our previous C+C measurements \cite{hades_cc1gev,hades_cc2gev},
there is a strong excess over the cocktail of known long-lived sources.
We know furthermore from our comparison \cite{hades_nn} of dielectron
production in free nucleon-nucleon and C+C reactions that in the light carbon-carbon
system the excess yield $N_{exc}$ is of baryonic origin: $\Delta$ decays and
$NN$ -- i.e.\ mostly $pn$ -- bremsstrahlung.  The C+C reaction seems to be
in first order an incoherent superposition of elementary $NN$ processes.
In addition, whereas between 1~and 2\agev\ $\eta$ production increases
steeply with bombarding energy (from sub-threshold to above threshold),
the excess yield rises like pion production, i.e.\ only mildly.  This means
that our isospin-averaged $\frac{1}{2} [pp+np]$ pair spectrum measured at
1.25~\gev\ \cite{hades_nn} can serve as $NN$ reference for the
$\eta$-subtracted pair yield observed in the present 1.76\agev\ Ar+KCl
run as well, both normalized to their respective $\pi^0$ multiplicity.
Note, however, that this reference is of course available only up to its
kinematic cutoff at $M_{ee} = 0.55$~\gevcc, corresponding to the 1.25~\gev\
bombarding energy used in the $NN$ experiments.


\begin{figure}[!hbt]
  \mbox{ \epsfig{width=0.90\linewidth, figure=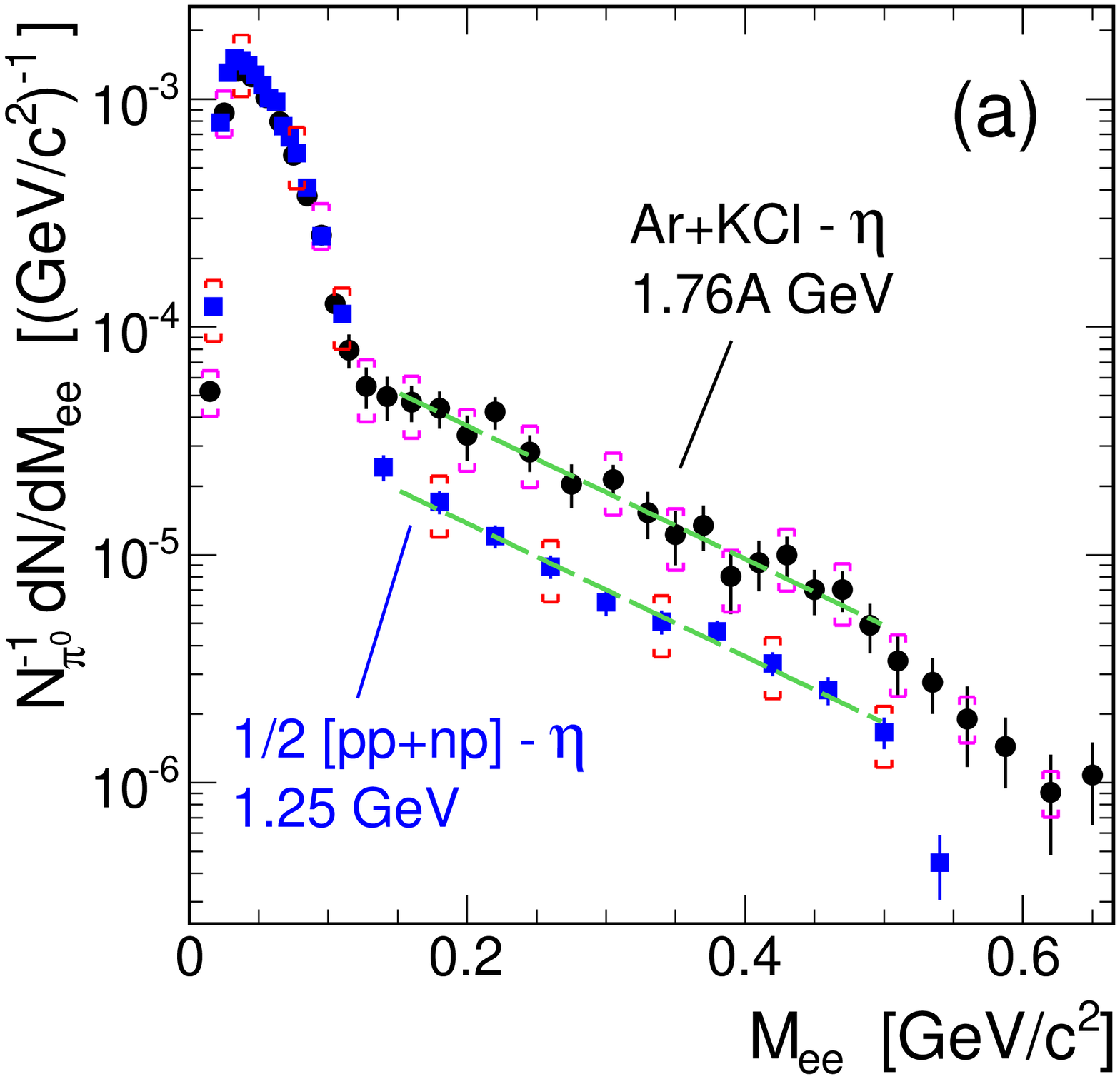}}
  \quad
  \mbox{ \epsfig{width=0.90\linewidth, figure=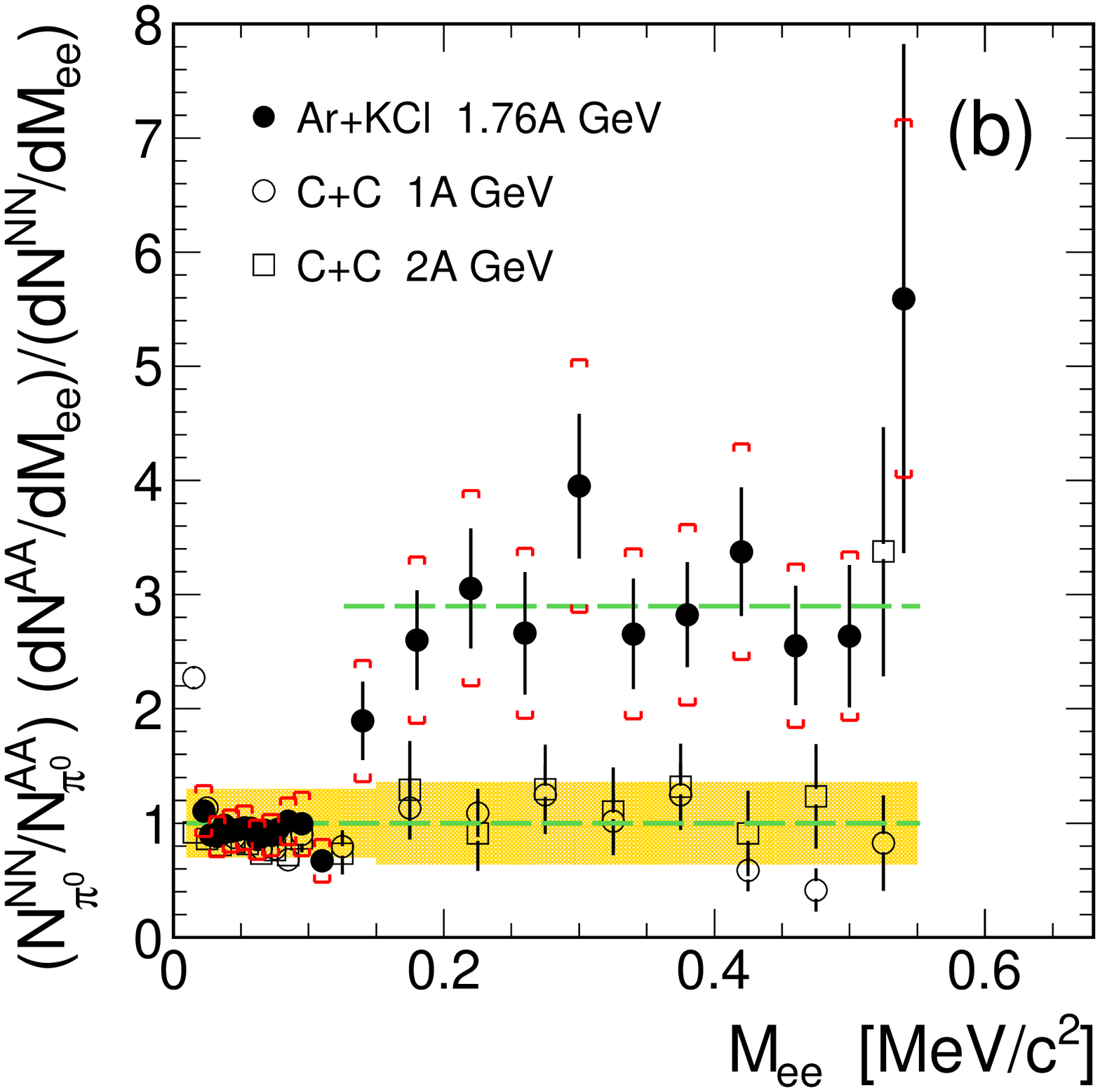}}
  \vspace*{-0.2cm}
\caption{(Color online)
         {\bf (a)} Comparison of the Ar+KCl invariant-mass distribution with an
         isospin-averaged reference from \pp\ and \np\ data \cite{hades_nn}.  For
         clarity systematic error bars are shown only on every second data point
         (vertical bars are statistical, cups are systematic). 
         Both data sets are normalized to their respective
         pion multiplicity and have their respective $\eta$ Dalitz yield
         subtracted.  The dashed lines are meant to guide the eye.
         {\bf (b)} Ratio of the heavy-ion mass distributions (Ar+KCl and
         C+C) to the 1/2 [pp+np] reference, whose total error (statistical
         and systematic added quadratically) is indicated by the shaded
         band.  Note that all data sets are shown within the acceptance of
         the Ar+KCl experiment.}
\label{NN_CC}
\end{figure}

Figure~\ref{NN_CC} (top) shows the Ar+KCl \ee\ invariant-mass distribution
after subtracting the simulated $\eta$ component and normalizing to
$N_{\pi^0}$, together with the $NN$ reference from \cite{hades_nn}
(adjusted to the acceptance, i.e.\ magnetic field and momentum cuts, of the
present experiment), also $\eta$ subtracted and normalized to its $\pi^0$
multiplicity (averaged from p+p and p+n data).  In this comparison we
do not correct for the slight isospin asymmetry of the Ar+KCl system ($N/Z = 1.15$).
Due to the normalization and the uae of a common acceptance both distributions agree
in the $\pi^0$ Dalitz peak.  They differ, however, strikingly for masses between 0.15
and 0.5 \gevcc\ where the yield from the heavy system exceeds the $NN$
reference by a factor of $\simeq 2.5-3$.  This is also
visible in the lower part of Fig.~\ref{NN_CC} where the ratios of the
following pair yields are shown: Ar+KCl/\NN, and C+C/\NN\ for 1 and 2\agev.
For this the C+C data were taken from \cite{hades_cc1gev,hades_cc2gev}
and transformed into the acceptance of the present experiment.
The Ar+KCl/\NN\ ratio is very close to unity at low masses, dominated by
the $\pi^0$ Dalitz pairs, but for $M>0.15$ \gevcc\ it rises to about 3,
indicating the onset of processes not accounted for in the reference system.
Both representations prove that a qualitative change happens in the nature
of the excess yield when going to the heavier system.  Consequently, in
contrast to the C+C system, Ar+KCl can not anymore be seen as a
superposition of independent \NN\ collisions.  A more complex picture involving
multi-body and multi-step processes and maybe even in-medium modifications
of the involved hadrons is required.  Note also that a scaling with the
number of binary nucleon-nucleon collisions $N_{coll}$ might be more
appropriate to describe the observed variation of the excess yield with
system size.  Indeed, $\langle N_{coll} \rangle$ calculated within a
Glauber approach \cite{glauber} increases faster than
$\langle A_{part} \rangle$ when going from our LVL1 C+C to LVL1
Ar+KCl events, namely by a factor 6.1 for $\langle N_{coll} \rangle$ vs.\ 4.5
for $\langle A_{part} \rangle$.

\begin{figure}[!ht]
  \mbox{\epsfig{width=0.90\linewidth, figure=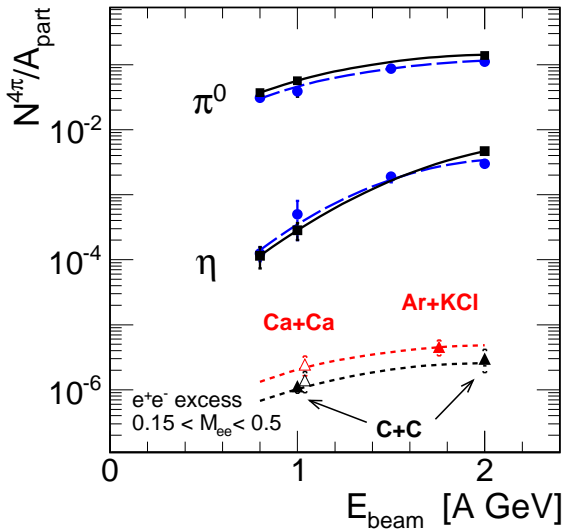}}
  \vspace*{-0.2cm}

  \caption{(Color online)
   Inclusive multiplicity per participant, $N^{4\pi}/\langle A_{part} \rangle$,
   as function of beam energy of the \ee\ pair excess over the $\eta$
   Dalitz yield, and of $\pi^0$ and $\eta$ production in heavy-ion reactions.
   The excess yield, defined in the mass range \mee$=0.15 - 0.5$~\gevcc and
   extrapolated to $4\pi$, is shown for HADES Ar+KCl and C+C data
   (full triangles) \cite{hades_cc1gev,hades_cc2gev}
   as well as for DLS data (open triangles) \cite{dls_prl_porter}.
   The $\pi^{0}$ and $\eta$ results are from TAPS measurements in
   C+C (squares, solid curves) and Ca+Ca (circles, long-dashed curves)
   collisions \cite{taps_CC,taps_thermal}.  The curves are polynomial
   fits to these data used to interpolate the multiplicities as a function
   of bombarding energy (see \cite{taps_CaCa}).  For easier visual
   comparison with the energy dependence of the dielectron excess
   the latter is overlayed with the $\pi^{0}$ curves (short-dashed)
   down-scaled by factors of $1.8\cdot10^{-5}$ for C+C and
   $4.3\cdot10^{-5}$ for Ar+KCl.
   }

   \label{excess_systematics}
   \vspace*{-0.0cm}
\end{figure}

Combining the dielectron results from HADES and from the former DLS experiment
we can now study the evolution of the excess over cocktail with beam energy {\em and}
system size.  To do so we have compiled in Fig.~\ref{excess_systematics}
the excess yields integrated over the mass region $M_{ee} = 0.15 - 0.5$ \gevcc\
from all available reaction systems \cite{dls_prl_porter,hades_cc1gev,hades_cc2gev}.
For comparison, inclusive $\pi^0$ and $\eta$ multiplicities measured in photon
calorimetry with the TAPS detector \cite{taps_CC,taps_CaCa} are plotted as well.
Note that all yields are extrapolated to the full solid
angle\footnote{Assuming similar geometric acceptances for excess pairs and $\eta$ Dalitz pairs.}
and are normalized to their respective average $A_{part}$ in order to
compensate for differences in the centrality selection of the various experiments.
The normalization also takes out the trivial system-size dependence of the
yields, as visible from the closeness of the C+C and
Ca+Ca meson curves\footnote{We consider here the systems Ar+KCl
and Ca+Ca as being equivalent in size and isospin.}.
The somewhat smaller pion multiplicity per $A_{part}$ of Ca+Ca can be
attributed to meson re-absorption in this larger system.
Note, however, that the eta multiplicities start out with the opposite behavior
at low beam energy and switch only around $E_{beam} = $1.5\agev\ to the
absorption-dominated scaling.  This crossing can be explained by the
transition from the sub-threshold regime, where multi-step processes
favored by a larger reaction volume are important \cite{taps_AuAu}, to above
threshold production.  

Next one can see that the dielectron excess follows pion production
with rising bombarding energy, as we stated already before \cite{hades_cc1gev}.
This turns out to be true for both the C+C and Ca+Ca collision
systems, as one can see from the excellent match with the scaled-down pion
production curves in the figure.  Such a behavior has been interpreted
as being characteristic of a production mechanism not driven by the
excitation of heavy resonances, but rather by low-energy processes
like pion production and propagation involving the $\Delta$ and, maybe,
low-mass $\rho$ excitations as well as bremsstrahlung \cite{hades_nn}.

As already pointed out, the systematics of excess yields has become
sufficiently rich to allow also for a study of the system-size dependence of
the electromagnetic radiation from the nuclear medium.  Above we concluded that
the comparison of the excess yield obtained in Ar+KCl with the $NN$
reference reveals a non-trivial behavior of pair production.  This is also
supported by Fig.~\ref{excess_systematics} where an increase of a factor
$\simeq2$ is visible when moving from the C+C to the Ca+Ca system.
Evidently the excess yield must scale faster than linear with $A_{part}$
in contrast to the behavior of e.g.\ pion production in heavy-ion reactions.
The mass dependence can indeed be further quantified by adjusting a
$N_{exc} \propto \; A_{part}^{\alpha}$ scaling law to the yields.  Using
the Ar+KCl excess obtained at 1.76\agev\ and interpolating the C+C
excess for that beam energy we get a coefficient $\alpha = 1.41^{+0.19}_{-0.27}$.
Using instead the DLS point measured at 1.04\agev\ results in a similar
scaling coefficient.  Note that, when varying $\langle A_{part} \rangle$
by comparing systems of different size, the corresponding scaling exponent
for pion production has been found to be $\alpha \simeq0.85$ \cite{taps_thermal,Reisdorf},
independent of beam energy, and the one for eta production to be $\alpha \simeq1.2$
at 1\agev, decreasing to $\alpha \simeq0.8$ at 2\agev\ \cite{taps_thermal}.
These numbers confirm that the dielectron excess scales with system size very
differently than the freeze-out yields of pions and eta mesons.

All of our observations put together may be interpreted as the onset
multi-body and multi-step processes in the hot and dense phase created
in collisions of nuclei of such a size.
The penetrating nature of the dilepton probe offers then
a natural explanation for the behavior of the excess \ee\ yield
if one keeps in mind that for a sufficiently large number of
participating nucleons or, in other terms, for a sufficiently
large collision volume the detected radiation is the integral over the full
time of the complex heavy-ion reaction and not just a snapshot at
freeze-out.  It will be interesting to follow up on this trend when
still heavier systems are added to the systematics.

\subsection{High-mass pairs from omega decays}

\begin{figure}[!hb]
  \mbox{\epsfig{width=0.90\linewidth, figure=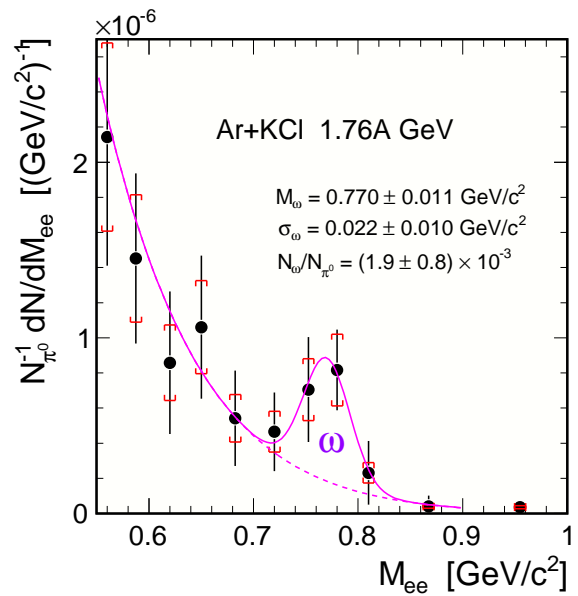}}
  \vspace*{-0.2cm}
  \caption{(Color online) Zoom into the measured \ee\ mass distribution in Ar+KCl
   (efficiency corrected and normalized), together with a least-squares
   fit of the $\omega$ vector meson peak by a sum of
   a Gauss function and an exponential background (see text for details).
   Error bars are like in Fig.~\ref{av_mass_spectrum}.}
   \label{omega_peak}
   \vspace*{0.0cm}
\end{figure}

We turn now to the high-mass region of the invariant mass spectrum and, in
particular, to the clearly protruding peak structure which we attribute
to the direct decay, $\omega \rightarrow e^+e^-$, of the omega vector
meson.  A linear zoom-in onto the vector-meson region is shown in
Fig.~\ref{omega_peak}.  The peak visible at the $\omega$ pole position
holds about 40 reconstructed pairs, limiting unfortunately the extent to
which one can possibly go in its analysis.  These data constitute the
very first observation of omega production in a heavy-ion reaction at
such a low beam energy, in fact, an energy even below the production
threshold in free \NN\ collisions ($E_{thr}^{NN} = 1.89$\agev).
One expects that most of the omegas contributing to this peak decayed
after having left the reaction zone, i.e.\ after freeze-out.  Recently
measured $\omega$ photoproduction cross
sections \cite{taps_omega1,taps_omega2,taps_omega3}
have been interpreted \cite{LeupoldMetagMosel} in the sense
of a strong broadening (up to 150~\mev) of the decay width of this meson
in the nuclear medium already at normal nuclear density.
We do not observe such a modification in our omega signal: the shape of
the observed peak is solely determined by the detector response, i.e.\
by the intrinsic momentum resolution of the HADES tracking system.
In this mass region also $\rho^0$ decays and baryonic resonance decays
are expected to contribute to the dielectron yield, but they
add up to a broad continuum underneath the omega peak.
For masses above 0.9 \gevcc\ the statistics is running out quickly and
there is no recognizable structure at the pole position of the $\phi$
meson ($M_{\phi} = 1.019$ \gevcc).

All of this justifies fitting the whole mass region with the sum of a Gauss
shape and an exponential function, as shown in Fig.~\ref{omega_peak}.
The fit ($\chi^2/ndf = 11.8/18$) provides a peak position
of $M_{\omega} = 0.770 \pm 0.011$ \gevcc, a width
of $\sigma_{\omega} = 0.022 \pm 0.010$ \gevcc, and an integrated signal over the
continuum corresponding to $(3.9 \pm 1.7) \cdot 10^{-8}$.   The peak centroid
agrees hence within about one standard deviation with the listed $\omega$
pole position at 0.783 \gevcc\ \cite{pdg2010}.
Furthermore, detector simulations show that part of the observed down-shift
($\simeq10$~\mev) is due to the combined energy loss of the electron
and positron in the inner part of the HADES detector.  The peak width is
dominated by the HADES mass resolution $\sigma/M$ at the
$\omega$ pole mass of 3\%.  Finally, its integral has been corrected for
the branching ratio of the direct \ee\ decay \cite{pdg2010} as well as for
the acceptance of 0.29 (obtained from a Pluto simulation done for a thermal
source with a temperature of $T=84\pm2$~\mev, as found in our
$\phi \rightarrow$ K$^+$K$^-$ analysis \cite{hades_phi} for the $\phi$ meson).
This resulted in a normalized yield of
$N_{\omega}/N_{\pi^0} = (1.9 \pm 0.8) \cdot 10^{-3}$, corresponding to an
omega LVL1 multiplicity of $M^{LVL1}_{\omega} = (6.7 \pm 2.8) \cdot 10^{-3}$.
Fits with more sophisticated peak shapes taking into account the slightly
asymmetric momentum response of the detector gave very similar results.
The acceptance correction depends mildly on the phase-space distribution
used in the Pluto simulation:  it ranges from 0.34 at $T=50$~\mev\ to 0.24 at
$T=140$~\mev\ (see also the discussion of the pair $m_{\perp}$ slopes in the
next subsection).  It depends even less on the assumed polar distribution:
5\% decrease when varying $A_2$ from 0 to 1.  All those effects are finally
subsumed into an additional systematic error on the multiplicity of 25\%.
With the $\omega$ yield known, both its contributions -- Dalitz and direct --
to the pair cocktail can be simulated in Pluto; they are shown together
with the mass spectrum in Fig.~\ref{av_mass_spectrum}.  The $\omega$ decays
contribute evidently only a small part to the total pair yield at
intermediate and low masses.  Note finally that the average $\omega$
momentum in the nucleus-nucleus center-of-mass within the HADES acceptance is found
from our data to be $p=0.43$~\gevc.  This is at least a factor two smaller
than the momenta typically observed in $\omega$ photoproduction
experiments \cite{taps_omega1,taps_omega2,taps_omega3}.

\begin{figure}[!hb]
  \mbox{\epsfig{width=0.90\linewidth, figure=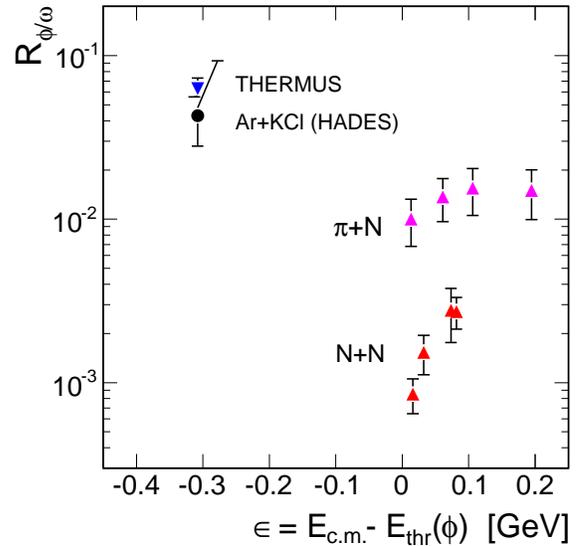}}
  \vspace*{-0.2cm}

  \caption{(Color online)
   Comparison of the $R_{\phi/\omega}$ ratio obtained in this work with
   its statistical model (THERMUS fit) value as well as with a compilation
   of data from elementary p+p and $\pi$+N reactions (see text).
   The ratio is plotted as a function of the excess energy $\epsilon$
   in the NN $\rightarrow$ NN$\phi$ and the $\pi$N$ \rightarrow$N$\phi$
   reactions, respectively.
   }
   \label{thermal_OZI}
   \vspace*{0.0cm}
\end{figure}

The $\omega$ multiplicity can be discussed in the context of either a
scenario of complete thermalization at freeze-out or, in the other extreme,
of production in elementary \NN\ collisions.  As HADES is a general-purpose
charged-particle detector, besides the dielectron results presented here,
a wealth of information has been obtained as well on hadron production in
Ar+KCl.  These findings have already been published in \cite{hades_pi}
on $\pi^{\pm}$, in \cite{hades_phi} on $K^+, K^-$, and $\phi$,
in \cite{hades_k0s} on $K^0_s$, in \cite{hades_Xi} on $\Xi^-$, and
finally in \cite{hades_Lambda} on $\Lambda$ and $\Sigma^{\pm}$.

In particular, from our $K^+ - K^-$ correlation analysis \cite{hades_phi},
a LVL1 $\phi$ multiplicity of
M$_{\phi} = (2.6 \pm 0.7(stat) \pm 0.1(sys)) \cdot 10^{-4}$ has
been found as well as a transverse-mass slope at mid-rapidity of
$T_{\phi} = 84 \pm 8$ \mev.  Together with the $\omega$ multiplicity, this
gives a $\phi/\omega$ ratio of
$R_{\phi/\omega} = 0.043^{+0.050}_{-0.015}(stat) \pm 0.011 (sys)$.
The experimental ratio can be compared to various predictions, running from
pure $m_{\perp}$ scaling in $4\pi$ solid angle, giving $R \simeq 0.042$, to a
full-fledged statistical hadronization model calculation performed with the
THERMUS code \cite{thermus} fitted to our hadron yields \cite{hades_Lambda}
and resulting in $R = 0.063 \pm 0.008$.  Hence, statistical
descriptions agree within error bars with the experimental $R_{\phi/\omega}$.
As already discussed in ref.~\cite{hades_Lambda}, the THERMUS model
does well in reproducing our measured hadron yields, including those of
particles with open or hidden strangeness, with the notable exception
of the double-strange $\Xi^-$ which, however, at 1.76\agev\ is produced
far below its threshold of 3.57~\gev\ in free $NN$ collisions.

The opposite extreme to complete thermalization is given by elementary \NN\
and $\pi + N$ reactions where the $\phi/\omega$ ratio is traditionally investigated
in the context of the so-called OZI rule violation \cite{OZI_review,OZI_NN,OZI_piN}.
The ratio obtained in those reactions for small values of the excess energy
($\epsilon = E_{c.m.} - E_{thr}$) notoriously exceeds predictions based on
the $\phi - \omega$ mixing angle and is sometimes related to a possible
$s \overline{s}$ admixture in the nucleon ground-state wave
function.  Figure~\ref{thermal_OZI} shows $R_{\phi/\omega}$ obtained in
this work and the THERMUS value from a fit to HADES data together with
results from elementary \pp\ \cite{disto,anke1} and $\pi+N$ \cite{baldini}
reactions, all plotted as function of the excess energy in the
NN$\rightarrow$NN$\phi$ and $\pi$N$\rightarrow$N$\phi$ reactions, respectively.
This is different from the common definition in literature where the
$\phi$ and $\omega$ yields are both taken at the {\em same} excess energy,
corresponding hence to different bombarding energies, whereas we take the
ratio of yields measured at a common beam energy.  In fact, to do this,
we divided the measured $\phi$ cross sections by an interpolation of the
omega cross sections based on the parameterization proposed in \cite{OZI_NN}
and, in case of the \pp\ data, updated in \cite{cosytof}.
One can see from the comparison that in the heavy-ion reaction the ratio
$R_{\phi/\omega}$ is more than an order of magnitude larger than in $NN$
collisions and also at least a factor 3 - 5 larger
than in pion-induced processes.  One should furthermore keep in mind
that mostly low-momentum pions are produced in 1 -- 2\agev\ heavy-ion
reactions while the $\phi$ production threshold is at $p_{\pi}$ = 1.56~\gevc;
in \NN\ collisions the production threshold is at 2.60~\gev.
Consequently, the $\phi$ meson is produced sub-threshold here ($\epsilon < 0$)
and more complex, multi-step processes involving short-lived
resonances \cite{BUU_phi} and/or hyperons \cite{catalytic_phi} might contribute.
On the other hand, the ratio could also be influenced by final-state
effects of the vector mesons in the nuclear medium
(see e.g.\ \cite{LeupoldMetagMosel} for a discussion of $\omega$
and $\phi$ absorption).  In the end, our observation seems to support the
picture of meson production in a rather long-lived and thermalized fireball.

\subsection{Dielectron $m_{\perp}$ distributions}

We present now phase-space distributions of \ee\ pairs in Ar+KCl.
When discussing the pair mass spectrum (see Fig.~\ref{av_mass_spectrum})
we distinguished different mass regions of interest.  Indeed, the pair spectrum
is a complicated cocktail emitted from various processes and at different phases
of the heavy-ion reaction.  As pointed out above, at low masses the
situation is rather simple because this region is dominated by pairs from
the $\pi^0$ Dalitz decay.  In all other regions, however, multiple sources
contribute and their disentanglement is not trivial.
We have fortunately constraints on the $\eta$ Dalitz yield from earlier
TAPS measurements and now also on the $\omega$ Dalitz yield from our own
analysis of the multiplicity of this particle (see discussion in III.C).

\begin{figure}[!hb]
    \mbox{\epsfig{width=0.9\linewidth, clip=true, figure=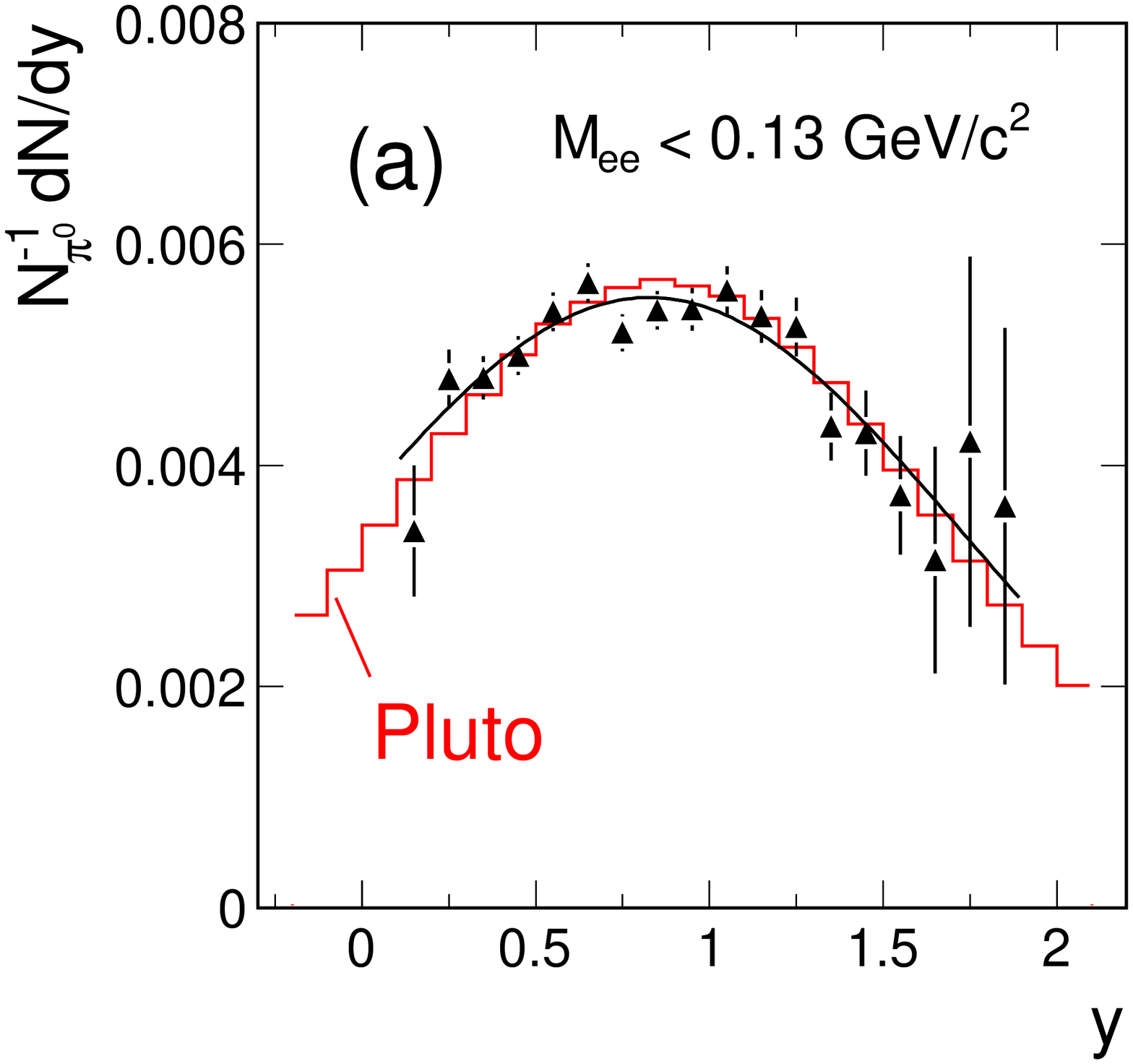}}
    \quad
    \mbox{\epsfig{width=0.9\linewidth, clip=false, figure=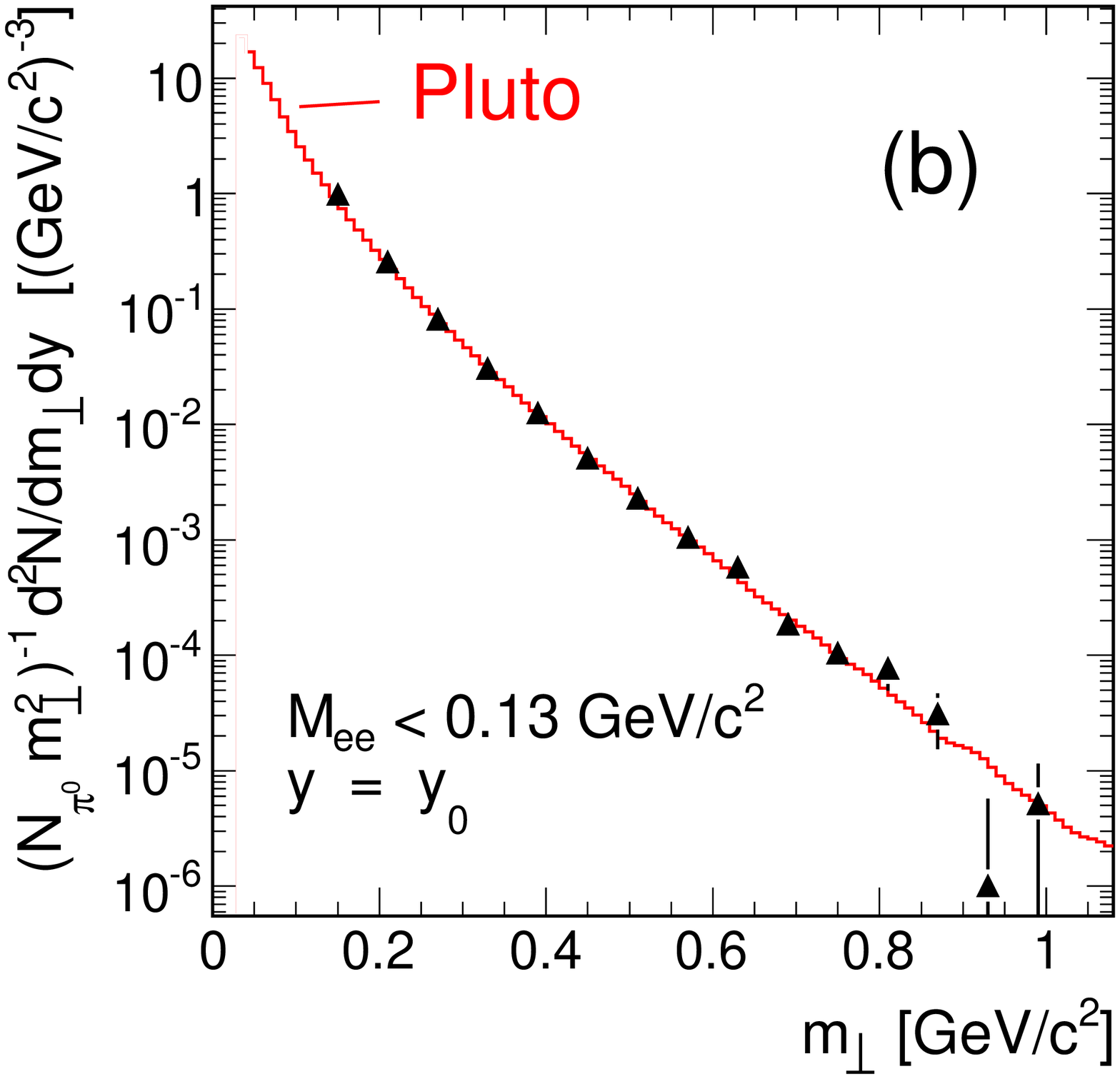}}
    \vspace*{-0.2cm}
\caption{(Color online) Reconstructed rapidity (a) and transverse mass at mid-rapidity
     (b) distributions of \ee\ pairs with $M_{ee} < 0.13$ \gevcc.
     Data are normalized to $N_{\pi^0}$ as well as corrected with the
     detector efficiency and acceptance.  The error bars are statistical.
     The histograms correspond to a simulated (Pluto) cocktail of thermal sources.
     A Gauss fit to $dN/dy$ (top, solid curve), shown as well, results in
     $\langle y \rangle = 0.83\pm0.03$ and $\sigma_y = 0.91\pm0.07$.
     }
\label{lowmassdist}
\vspace*{0.0cm}
\end{figure}

To characterize the dielectron yield beyond its mass distribution one has to
reconstruct other pair observables, in particular its longitudinal and
transverse phase-space population.  The longitudinal dimension is usually
covered by plotting rapidity density as function of rapidity, $dN/dy$, and
the transverse one by plotting either $dN/dp_{\perp}$ or $dN/dm_{\perp}$
with $m_{\perp} = \sqrt{p_{\perp}^2 + M_{ee}^2}$ as function of $p_{\perp}$
or $m_{\perp}$.  The thermal nature of a particle source can be best
recognized by plotting either its $1/m_{\perp}^2 \; dN/dm_{\perp}$
distribution at $y=y_0$ or its $1/m_{\perp}^{3/2} \; dN/dm_{\perp}$
distribution integrated over all rapidities \cite{hagedorn}.
Indeed, in semi-logarithmic representation and
for $m_{\perp} >> T$, both functions turn into a straight line where the
inverse-slope parameter $T$ may be interpreted as the source temperature.
In case of measuring dilepton pairs the situation is further complicated
by the decay kinematics of the three-body Dalitz decays.  Consequently,
the exact nature of the parent distribution can be distorted in the
observed \ee\ distribution.  Note that, in order to obtain meaningful
slopes, these distributions have to be corrected not only for efficiency
but also for acceptance including the detector geometry as well as
momentum and opening angle cuts.  As mentioned in the discussion
of the $\omega$ multiplicity determination, the acceptance correction
has been obtained from Pluto simulations of a full pair cocktail
(with the source parameters listed in Tab. \ref{pluto_params}),
while varying its source parameters to quantify systematic effects.
The pair acceptance has thereby been determined as a one-dimensional
function of transverse mass, averaged over rapidity within a given
mass bin, and vice-versa.  We have verified that this procedure
gives results compatible with the more complex multi-dimensional
correction as function of mass, transverse momentum, and rapidity.
The resulting normalized $dN/dy$ and $dN/dm_{\perp}$
spectra are shown in Fig.~\ref{lowmassdist} for pairs of $M_{ee} < 0.13$ \gevcc,
together with the corresponding simulated spectra.  This low-mass bin
-- being dominated by $\pi^0$ Dalitz yield -- is described to better than
10\% by our Pluto event generator, as seen from the overlayed histograms. 
This agreement further strengthens our confidence in the pion phase-space
distribution used in the simulation as well as in our dielectron reconstruction
procedures in general.

\begin{figure}[!ht]
\centering
\mbox{\epsfig{width=0.9\linewidth, clip=true, figure=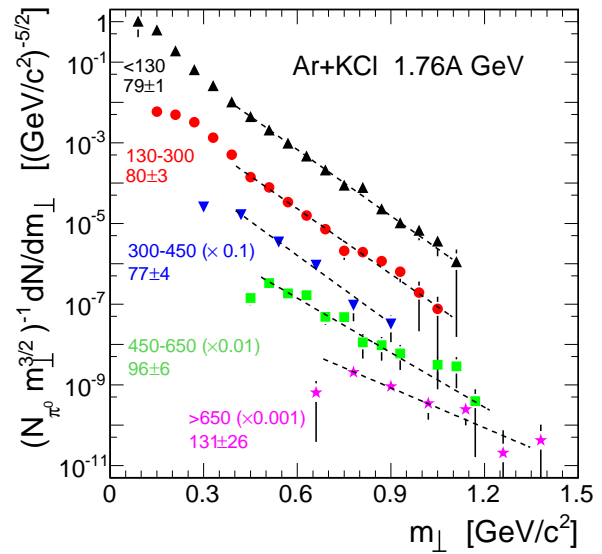}}
\vspace*{-0.2cm}
\caption{(Color online) Reconstructed pair $1/m_{\perp}^{3/2} \; dN/dm_{\perp}$
     distributions, normalized to the $\pi^0$ multiplicity, for the full
     rapidity range and different mass selections given
     in the l.h.s.\ legends (in \mev).  Efficiency and acceptance
     corrections are applied; error bars are statistical.
     Exponential fits to the high-$m_{\perp}$ region of the data are shown
     as dashed curves with the corresponding inverse-slope parameter given
     (in \mev) in the second line of the legends.
     Note also the scaling factors (in parentheses).  
     }
\label{dilepton_slopes}
\vspace*{0.0cm}
\end{figure}

In the context of this analysis we have also done a careful
investigation of the signal purity, as one might fear that
particularly the high $m_{\perp}$ pairs could be contaminated by
misidentified high-momentum hadron tracks and/or fake tracks.  This
purity study has been done with an event mixing technique and confirmed as
well with full simulations of the reconstruction and particle identification.
Indeed, while our lepton purity is on average better than 0.95, it decreases
with increasing lepton momentum, resulting nonetheless in a dielectron purity
which remains better than 0.7 up to $m_{\perp}$ values of 1.5~\gevcc.  Note
that hadron and fake impurities in the lepton sample lead to uncorrelated
pairs only, and thus increase the combinatorial background which is of
course subtracted, as discussed in section~II.  We have checked in
simulations that the CB subtraction indeed removes these additional
uncorrelated contributions.

To take advantage of our full pair statistics, we have opted to use the
$1/m_{\perp}^{3/2} \; dN/dm_{\perp} = N_{\circ} \, exp(-m_{\perp}/T)$ representation
in our systematic investigation of the transverse momentum distribution
for several bins of pair mass displayed in Fig.~\ref{dilepton_slopes}.
With increasing pair masses contributions from $\eta$ Dalitz,
$\Delta$ (and $N^*$) Dalitz, bremsstrahlung, and finally $\omega$, $\rho^0$
and (very few) $\phi$ decays are successively probed.
Although the limited statistics of our data required rather wide bins,
particularly for the highest masses, one can see a distinctive pattern emerge:
as one progresses from low to higher $M_{ee}$, the slope of the pair
transverse-mass spectra first remains approximately constant at $T$
around 80~\mev, but when approaching the vector meson region, it rises
steeply to reach a value as high as 130~\mev.

\begin{figure*}[!hbt]
\mbox{\epsfig{width=0.95\linewidth, clip=true, figure=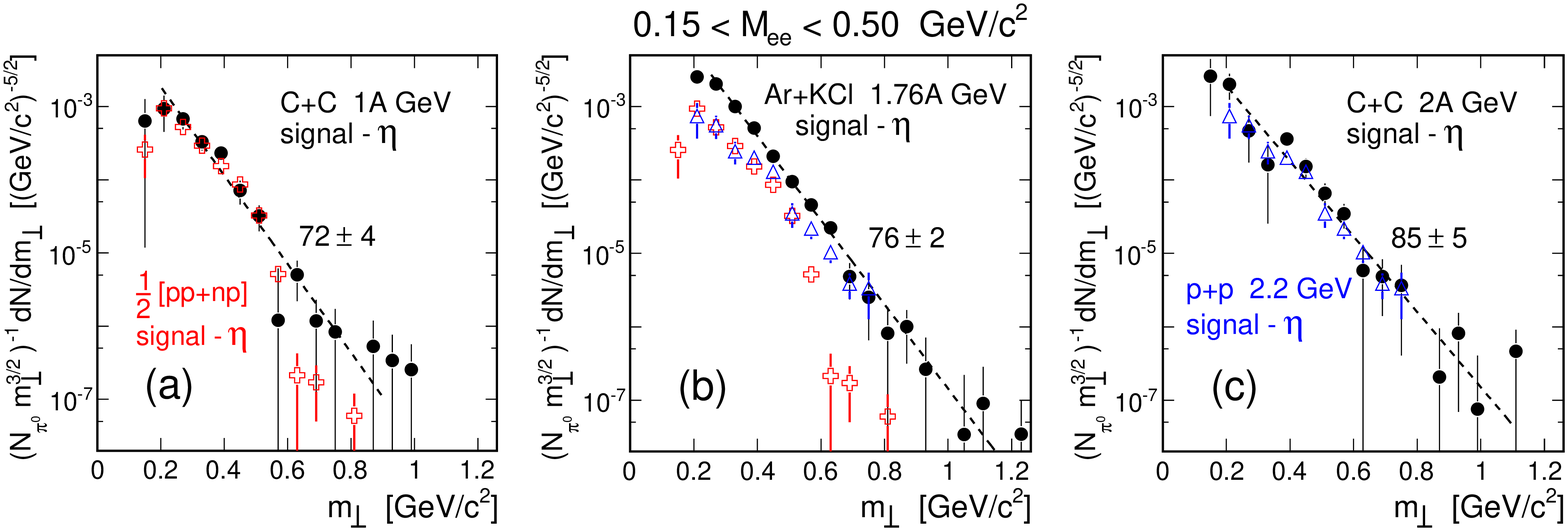}}
\vspace*{-0.2cm}
\caption{(Color online) Reconstructed $1/m_{\perp}^{3/2} \; dN/dm_{\perp}$
     distributions of pairs with $0.15<M_{ee}<0.5$~\gevcc\ and
     $-\infty<y<\infty$ in C+C at 1\agev\ (a),       
     in Ar+KCl at 1.76\agev\ (b), and in C+C at 2\agev\ (c).
     Reference spectra from elementary $NN$ collisions are also shown,
     namely the average of \pp\ and \np\ at 1.25 \gev\ (open crosses),
     and \pp\ at 2.2 \gev\ (open triangles). 
     Efficiency and acceptance corrections are applied.  All distributions have
     their respective $\eta$ contribution subtracted and are normalized to
     their respective pion multiplicity $N_{\pi^0}$.  Error bars are
     statistical.  Dashed lines are exponential fits, with the corresponding
     inverse slope parameter given in \mev.
     }
\label{midslope_systematics}
\vspace*{0.0cm}
\end{figure*}

While the first mass bin is dominated by $\pi^0$ Dalitz pairs, as emphasized
in discussing Fig.~\ref{lowmassdist}, the next two bins cover the
intermediate-mass region ($0.15 < M_{ee} < 0.5$~\gevcc), with contributions
from $\eta$ Dalitz, $\Delta$ Dalitz, $NN$ bremsstrahlung and maybe other sources.
This is the region of the pair excess which we would like to characterize
as much as possible.  To do this, we have again subtracted the eta
component simulated with Pluto by making use of the known $\eta$
multiplicity and source temperature.  The resulting excess $dN/dm_{\perp}$
distribution is shown in  Fig.~\ref{midslope_systematics} together with
corresponding data obtained in C+C at 1 and 2\agev, as well as with a
reference from elementary nucleon-nucleon collisions, obtained from the average
of our \pp\ and \np\ results at 1.25~\gev\ \cite{hades_nn} as discussed in III.B.
The spectrum from elementary \pp\ collisions at 2.2~\gev\ \cite{benjamin_thesis}
is also shown, but at this energy, unfortunately, the corresponding \np\
yields needed for isospin averaging are not available.  All distributions
are normalized to their respective neutral pion multiplicity, $N_{\pi^0}$,
and have their respective $\eta$~Dalitz contribution subtracted\footnote{For 2.2~\gev\ p+p only
  the exclusive $\eta$ production has been subtracted.}.  The figure shows
that -- within error bars and up to $m_{\perp} \simeq 0.5$, respectively
$\simeq 0.8$ -- the light C+C system behaves at both bombarding energies
very much like the $NN$ reference:  normalized yields and slopes agree to a
large extent.  Note again that the reference spectrum can have yield only
up to its kinematic cutoff at $m_{\perp} = 0.55$~\gevcc\
(0.89~\gevcc\ for 2.2~\gev\ p+p).  In contrast to C+C, however,
the Ar+KCl system displays a large excess over the elementary reference,
just as found already in the comparison of the pair invariant mass spectra.

\begin{figure}[!hb]
\mbox{\epsfig{width=0.90\linewidth, clip=true, figure=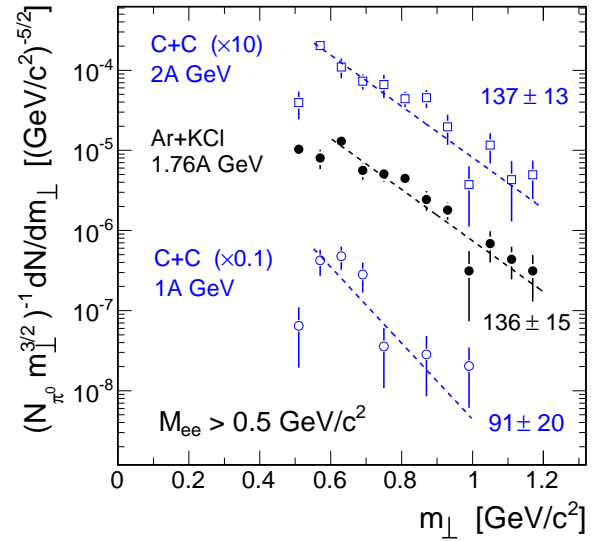}}
\vspace*{-0.2cm}
\caption{(Color online) Comparison of dielectron $1/m_{\perp}^{3/2} \; dN/dm_{\perp}$
     distributions with $M_{ee} > 0.5$~\gevcc and $-\infty<y<\infty$
     in Ar+KCl and C+C (see text for details).
     Efficiency and acceptance corrections are applied.  The distributions are
     normalized to the respective reaction pion yield $N_{\pi^0}$.  Error bars
     are statistical.  Dashed lines are exponential fits, with their
     corresponding inverse slope parameter given in \mev.
     Note also the scaling factors.
     }
\label{hislope_systematics}
\vspace*{0.0cm}
\end{figure}

Moving finally to the large kinetic slopes found in the two upper mass
bins, we note that this observation is surprising and difficult to
reconcile with the assumption of a completely thermalized particle source.
As one moves away from the three-body decays dominant below
the $\eta$ Dalitz edge at 0.547~\gevcc,
the two-body decays of the vector mesons contribute more and more, and
one expects indeed an increase of the slope parameter.  This is just a natural
consequence of the decay kinematics.  On the other hand, from our
THERMUS fit to the full set of hadron yields measured in Ar+KCl, we have
found a chemical freeze-out temperature of the fireball of 76~\mev\
(see section~III.C and \cite{hades_Lambda}) and kinetic slope parameters
in the range of 70 - 95~\mev\ \cite{hades_Lambda}, i.e.\ of similar
magnitude.  In particular, the slope at mid-rapidity of the $\phi$ meson
was found to be $84 \pm 8$~\mev\ in the $K^+ K^-$ channel \cite{hades_phi}.
Unfortunately our statistics is not large enough to allow a tight selection
around the $\omega$ pole mass and thus obtain the slope for a
clean $\omega$ sample.  From transport calculations \cite{filip_qm09}
we estimate that the pair cocktail selected by our uppermost mass window
(0.65 - 1.2 \gevcc) contains sizeable contributions from the vector mesons
$\rho, \omega$ and $\phi$, but its true composition remains of course uncertain.

Analyzing furthermore the corresponding $m_{\perp}$ slopes from the data sets
of the C+C system \cite{hades_cc1gev,hades_cc2gev}, and comparing them
with the present Ar+KCl result (see Fig.~\ref{hislope_systematics}),
we find a comparatively large slope in the light C+C system at 2\agev.
At the lower beam energy of 1\agev, however, the C+C slope is found
to be much smaller, but still large when\ compared with the 58~\mev\
slope of charged pions observed in this system \cite{hades_CCpions}.

Presently we can only speculate about various effects that can lead to
such a behavior of the transverse mass distributions.  For example,
collective effects, like radial flow, produce large effective
temperatures and, in fact, yield slopes increasing with particle mass.
However, such a trend is not visible in the systematics of kinetic slopes that
we observed in Ar+KCl \cite{hades_Lambda} and, moreover, in this rather
small system the radial flow is not expected to be important \cite{Reisdorf2}.
Another mechanism that has been proposed is linked to the final-state
interactions of the produced vector mesons.  Indeed, if these interactions
are strongly momentum-dependent, they can modify the spectral distributions.
This has also been proposed as an explanation for the depletion
of $\omega$ yield observed at low $p_{\perp}$ in In+In collisions
by the NA60 experiment at the CERN-SPS \cite{na60_thermal}.  Re-absorption
cross sections of the $\omega$ meson in cold nuclear matter have been
calculated in a OBE approach by the authors of ref. \cite{omega_abs}
and they found them indeed large and strongly momentum-dependent.
Inserting these cross sections into transport calculations, they
also predicted observable effects in the transverse-mass spectra of
$\omega$ produced in heavy-ion collisions.  Note that re-absorption of
the vector mesons is closely related to their collisional broadening
in the nuclear medium.  From recent measurements of the transparency ratios
in $\omega$ \cite{taps_omega2,CLAS_omega} and $\phi$ \cite{SPring8_phi}
photoproduction the collisional broadening of both mesons has been found to
be quite large (at $\rho = \rho_0$, $\Gamma_{\omega} = 100-150$~\mev),
although its exact momentum-dependence could not yet be sufficiently constrained
(see \cite{LeupoldMetagMosel} for a discussion).  Yet another factor that
might influence the shape of the pair transverse mass distribution are
the spectral functions of the various dilepton sources contributing.
In particular, any enhancement at large masses due to increased
form factors, as predicted e.g. for resonance decays within
vector-dominance models \cite{faessler,iachello}, could lead to spectral distortions
transcended in their characteristic $m_{\perp}$ slopes.  To conclude,
the interpretation of the large dielectron slope parameters observed
in our heavy-ion data remains challenging.

\subsection{Dielectron angular distributions}

Angular distributions of the emitted dielectrons constitute yet another
observable of interest.  Various emission angles can be reconstructed.
Here we focus on two specific ones: (i) the center-of-mass polar angle $\theta_{c.m.}$,
i.e.\ the angle between the direction of the virtual photon in the $A+A$
reference frame and the beam axis, and (ii) the so-called helicity angle $\alpha$,
i.e.\ the angle between the direction of the virtual photon in the
reference frame of the mother particle (e.g.\ $\pi^0, \eta, \Delta$ for
the three-body Dalitz decays and the fireball for the two-body direct decays)
and the direction of the electron (or positron) in the pair
frame.  This particular choice of the latter angle corresponds to its
definition in the so-called Jacob-Wick frame \cite{JacobWick,LamTung}. 
Technically it requires a double Lorentz transformation of the lepton
momenta: first from the laboratory frame into the parent particle frame and
second from there into the virtual photon frame.
Considering the decay kinematics, one can convince oneself that the polar
angle of the dielectron will have at least reminiscence of the polar emission
angle of the mother particle for three-body decays (i.e.\ Dalitz decays)
and be equal to it in case of a two-body decay
(i.e.\ vector meson $\rightarrow e^+e^-$).
Likewise, the reconstruction of the helicity angle is exact only for
two-body decays and approximate for three-body decays, because
the third product goes undetected in the inclusive \ee\ reconstruction
we did.  Nonetheless, our simulations show that these angular distributions
are not completely attenuated if one makes the approximation that the
decaying particle is at rest in the nucleus-nucleus center-of-mass frame
and thus information can be gained on the parent particle and its decay.
The amount of attenuation depends on the specific decay kinematics and
on the source temperature of the mother particle.

\begin{figure}[!ht]
\mbox{\epsfig{width=0.8\linewidth, clip=true, figure=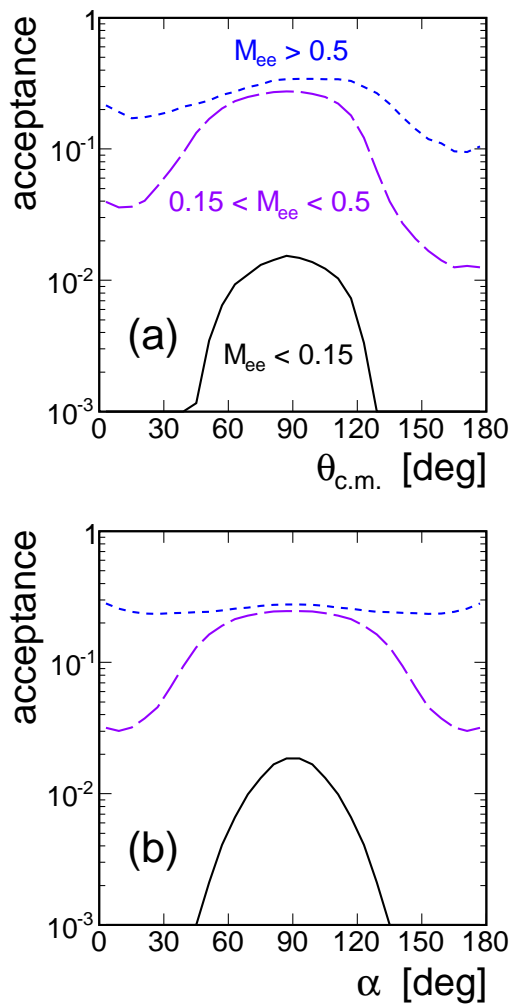}}
\vspace*{-0.5cm}
\caption{(Color online) Acceptance of the HADES detector for the dielectron
         nucleus-nucleus center-of-mass polar emission angle
         $\theta_{c.m.}$ (a) and helicity angle $\alpha$
         (b), represented for the three mass selections indicated in the figure.
     }
\label{helicity_acceptance}
\vspace*{0.0cm}
\end{figure}

\begin{figure*}[!htb]
\mbox{ \epsfig{width=0.35\linewidth, height=1.05\linewidth, figure=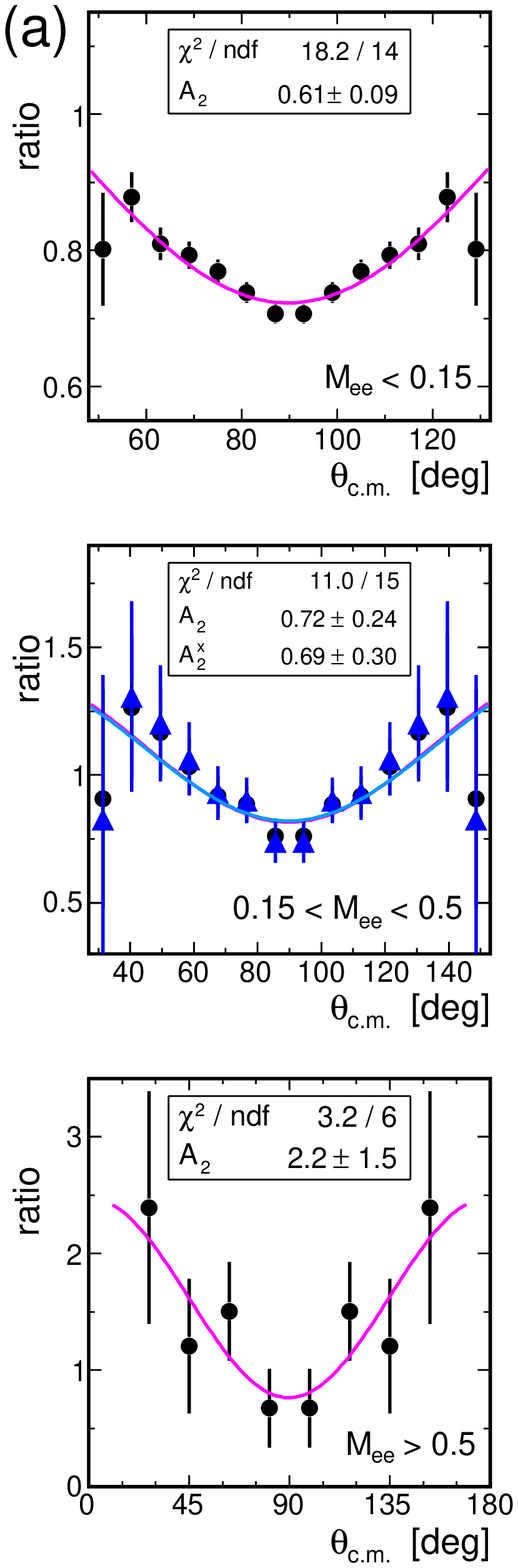}}
\mbox{ \epsfig{width=0.35\linewidth, height=1.05\linewidth, figure=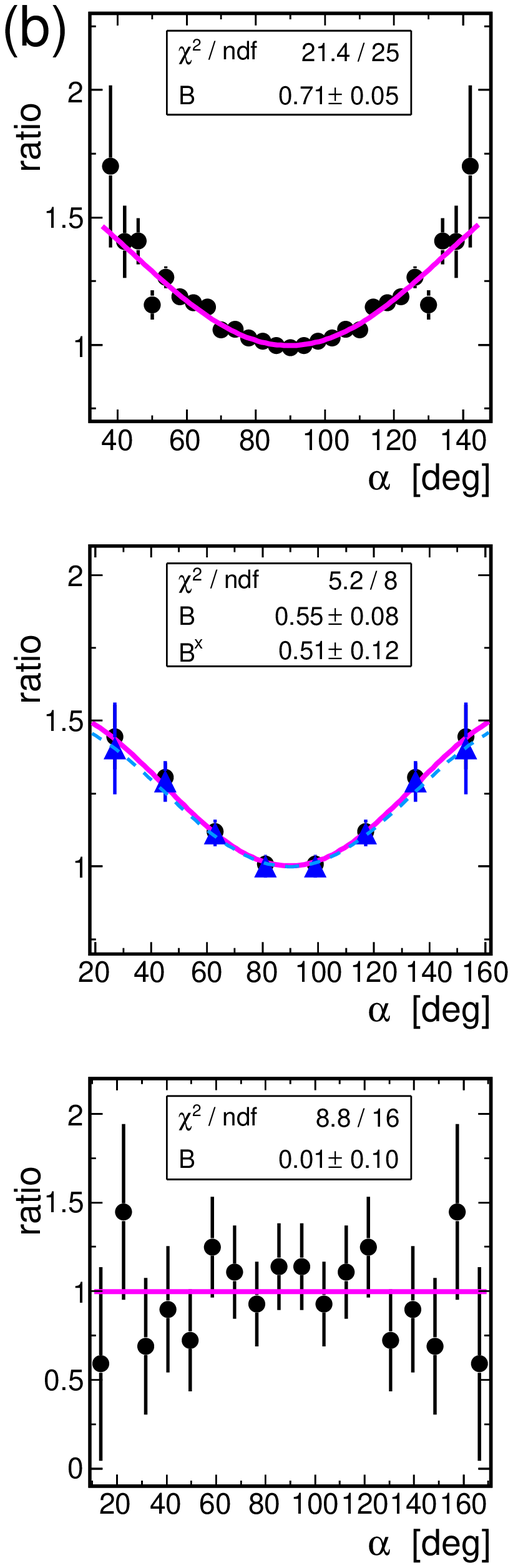}}

  \vspace*{-0.2cm}
  \caption{(Color online) Ratio of measured and simulated dielectron center-of-mass polar
   distributions $dN/d\theta_{c.m.}$ (a) and helicity distributions
   $dN/d\alpha$ (b) for three mass bins:
   $M_{ee}<0.15$, $0.15<M_{ee}<0.5$, and $M_{ee}>0.5$~\gevcc\ (from top to bottom).
   The error bars are statistical.  The Pluto cocktail simulation
   was done assuming isotropic emission and decay of the dileptons (see text).
   The curves are fits to the data yielding the anisotropy coefficients,
   i.e.\ polar $A_2$ and helicity $B$.  The coefficients $A^x_2$ and $B^x$
   result from fits (dashed curves) to the $\eta$-subtracted ratios
   (triangles).
  }

   \label{angular_distributions}
  \vspace*{0.0cm}
\end{figure*}

The helicity distribution is of particular interest as it allows to probe
the degree of polarization of the virtual photon.  It can be proven that
pseudoscalar Dalitz decays are self-polarizing \cite{kroll-wada,landsberg}
and lead to helicity distributions of the form $dN/d\alpha \propto 1 + B \cos^2{\alpha}$
with $B = 1$, where $\alpha$ is the helicity angle.
This expectation has been confirmed long ago for the $\pi^0$
in a study using the charge-exchange reaction $\pi^- p \rightarrow \pi^0 n$
\cite{samios} and again more recently in exclusive \pp\ measurements at 2.2~\gev\
performed by the HADES collaboration \cite{hades_helicity1}.  This HADES measurement
provided in addition the very first observation of the helicity
distribution in $\eta$ Dalitz decays.  The simulations we have done to determine
the sensitivity of our heavy-ion data to these effects reveal that the attenuation,
caused by the incomplete reconstruction of the Dalitz decays, reduces the helicity
anisotropy coefficient from unity to $B \simeq 0.6 - 0.7$ for
$\pi^0 \rightarrow \gamma e^+e^-$ and $\eta \rightarrow \gamma e^+e^-$
decays, i.e.\ leaving it still quite sizeable.
In case of the $\Delta \rightarrow N e^+e^-$ decay, the calculation of the
helicity distribution is much more involved, but it has been done and likewise
a distribution of type $1 + B \cos^2{\alpha}$ with $B \simeq 1$ is expected \cite{bratkovskaya_helicity1}.
Exclusive dielectron data taken with HADES in 1.25~\gev\ \pp\ collisions have also
confirmed the latter prediction \cite{hades_helicity2}.  Finally, in the two-body
decays -- $\rho^0, \omega, \phi \rightarrow e^+e^-$ -- the dilepton simply carries
the full polarization of the vector meson.  Measuring helicity angles might
hence help to unravel the different components of the pair cocktail
\cite{bratkovskaya_helicity1,bratkovskaya_helicity2,bratkovskaya_helicity3,
gulamov_helicity}.
These ideas have also been applied recently by the NA60 experiment to
characterize the thermal nature of high-mass dimuon radiation emitted
in high-energy heavy-ion reactions at the CERN-SPS \cite{na60_helicity}.

Experimentally the dielectron angular distributions are only obtained within the
detector acceptance shown in Fig.~\ref{helicity_acceptance}, which
they need to be corrected for.
We have done this by dividing a given reconstructed polar distribution
with a corresponding simulated cocktail distribution for which isotropic
emission of the parent particle was assumed.  Likewise, the experimental
helicity distributions were divided by the corresponding simulated
cocktail distribution assuming an isotropic emission of the decay lepton.
From our simulations we expect that such a ratio, besides correcting for the
acceptance, will reveal deviations of the data from isotropy.
The $\pi^0$ dominated low-mass pairs can again serve as a test bed for the procedure.
For these one expects to observe a polar distribution reminiscent of
the known pion polar anisotropy \cite{hades_pi} as well as the helicity
distribution typical for pseudoscalar Dalitz decays, although attenuated. 
Figure~\ref{angular_distributions} shows the ratio of the reconstructed/simulated
center-of-mass polar ($dN/d\theta_{c.m.}$) and helicity ($dN/d\alpha$)
distributions for three different pair mass bins.  Note that these ratios
have been reflected about $90^{\circ}$ and both halves added in
order to reduce statistical fluctuations.  The normalization is arbitrary.
The resulting angular distributions exhibit anisotropies which are
quantified by adjusting $1 + A_2 \cos^2{\theta_{c.m.}}$ and
$1 + B \cos^2{\alpha}$ forms, respectively.

The low-mass anisotropies are large and consistent with our expectations for
the neutral pion.  The fitted polar coefficient $A_2 = 0.61 \pm 0.09$
corresponds, according to our simulation, to an un-attenuated $A_2 = 0.76
\pm 0.11$, in agreement with the polar anisotropies of charged pions
observed in Ar+KCl \cite{hades_pi}, namely $A_2 = 0.75 \pm 0.05$.
The helicity, $B = 0.71 \pm 0.05$,
is attenuated by the thermal emission of the pion from its QED value $B = 1$,
again consistent with the expectation from our simulations.

Intermediate-mass pairs are more interesting because only about 25 - 30~\% of
their yield is exhausted by $\eta$ Dalitz pairs, the dominant excess part
being of non-trivial nature (see discussion in III.B).  Angular distributions
might provide some constraints on its possible composition.
The intermediate-mass bin in Fig.~\ref{angular_distributions} displays
large anisotropies as well, both for polar emission angles,
with $A_2 = 0.72 \pm 0.24$, and for helicity, with $B = 0.55 \pm 0.12$.
Taking into account the attenuation, this is again compatible with the
$1+\cos^2{\alpha}$ behavior typical for pseudoscalar meson, but also
$\Delta$ decays.  Subtracting the simulated $\eta$ contribution from the
data, the pure excess angular distributions have been obtained.  They are
represented as well in the figure, together with the corresponding fits, showing
that the anisotropies ($A_2^x = 0.69\pm0.30$ and $B^x = 0.51\pm0.17$) of
the excess yield turn out to be very similar to the ones of the $\eta$.
This suggests that a large fraction of the excess can be attributed to
decays of the $\Delta$ resonance for which indeed $B = 1$ is also expected.
One has to keep in mind, however, that the nucleon-nucleon bremsstrahlung
contributes as well in this mass region and has to be considered
in a full description.

The high-mass bin is unfortunately very low in statistics, but exhibits
nevertheless a strong  polar anisotropy ($A_2 = 2.2 \pm 1.5$),
whereas its helicity distribution is within its (large)
statistical errors compatible with $B = 0$.  This is to be expected,
if this mass bin contains large contributions from vector mesons emitted
from a completely thermalized source, just like it was observed in the
NA60 experiment \cite{na60_thermal}, although this is in conflict with the
observed large slope parameters.

\section{Conclusions and outlook}

The results on \ee\ production obtained with HADES in the medium-heavy \arkcl\
system at 1.756\agev\ show an intermediate-mass pair excess over long-lived
sources a factor 2--3 stronger than the one observed in elementary and \xcc\
reactions.  We have discussed the enhancement in the integral pair yields
as well as differentially in the pair mass and transverse mass
distributions.  We have argued that this behavior signals the onset of
influence of the nuclear medium on dilepton production.  We have
isolated the emission from the medium by subtracting the known
contributions from long-lived pair sources.  By presenting transverse
mass and angular distributions of the eta-subtracted yields, we have
been able to characterize this contribution further.  From the analysis
of the excess transverse-mass slope and angular anisotropies we concluded
that they are compatible largely with $\Delta$ Dalitz decays, suggestive
of resonance matter.

Furthermore, for the first time at SIS18 energies, a clear $\omega$ signal
could be observed in heavy-ion collisions, quantified and compared with
the prediction of a statistical hadronization model.  This result allows,
in particular, to put tight constraints on vector meson production
in heavy-ion collisions at beam energies of a few \gev.  From the
shape and position of the observed peak no direct indications could,
however, be found for medium modifications of this vector meson.
In fact, in case of the very strong broadening of the $\omega$ implied by
the interpretation of photoproduction data, our measurement would anyhow
have been sensitive only to the freeze-out, i.e. vacuum part of its yield.

A first and preliminary comparison of our Ar+KCl invariant mass
\ee\ spectrum with predictions of models based on transport theory has been
discussed in \cite{filip_qm09,martin_bormio10}.  While, both the Hadron String
Dynamics (HSD) model \cite{bratkovskaya_arkcl}
and the  Ultra-relativistic Quantum Molecular Dynamics (UrQMD) model \cite{UrQMD09}
achieve good agreement for invariant masses below 0.15~\gevcc, at
intermediate and higher masses the description of the pair yield
is not yet satisfactory.  Known reasons are the still imperfect description
of some of the elementary processes implemented in transport models,
namely bremsstrahlung and vector meson production \cite{hades_nn}, but also
the largely open question about how to treat possible in-medium modifications
of these processes.  On the other hand, these models suggest that the part
of our spectrum most sensitive to possible in-medium modifications should
be the region of excess yield, namely the dielectrons with masses of
0.5--0.8~\gevcc, which hence need to be characterized in detail.
With the additional and more differential \ee\ data presented in this
paper we have provided the information required to improve the theoretical
description of dilepton production in heavy-ion collisions.
Furthermore, once supplemented with data on yet heavier reaction systems,
these new results are expected to reveal and quantify medium
modifications of hadrons in warm and dense nuclear matter.
 

In summary, we have presented phase space distributions (invariant mass,
transverse mass, rapidity, polar angle, and helicity angle) of dielectrons
for the reaction Ar+KCl at 1.756\agev.  From these the following results
have been obtained: (i) observation of a strong excess (up to factor 3)
over \NN\ collisions of the pair yield at intermediate masses;
(ii) first observation of $\omega$ production in heavy-ion collisions
at such a low beam energy, yielding a $\phi / \omega$ ratio consistent
with maximal violation of OZI suppression; (iii) a systematic study of
transverse-mass distributions with the observation of unexpectedly
large inverse-slope parameters of up to 135~\mev, which might be
related to final-state processes and/or the spectral functions
of the contributing dilepton sources; and (iv) first exploitation
of the virtual photon polarization observable at low beam energies.  

Our studies on dielectron production in heavy-ion reactions will be pursued
over the next years with an upgraded HADES detector \cite{hades_upgrade}
which will have the ability to handle the high track densities from
truly heavy collisions, in particular also the \auau\ system.  These
data runs will thus provide the full systematics required to address
open questions on the origin and properties of the intermediate-mass
pair yield.  In parallel to this heavy-ion program, the HADES experiment
will also take up studies making use of the GSI secondary pion beams,
in elementary $\pi+p$ reactions as well as in $\pi+A$ reactions.
The pion-beam experiments will allow to conduct a comprehensive study
of the contribution of specific baryon resonances to dielectron emission,
in vacuum and in the nuclear medium.  The information gained this way
will in turn also help to improve our understanding of dilepton
radiation from the hot and dense hadronic medium produced in the
reactions with heavy ions.

\acknowledgments {
We thank R. Averbeck for providing valuable explanations on the TAPS data.
The HADES collaboration gratefully acknowledges the
support by the BMBF through grants 06MT9156, 06GI146I, 06FY91001,
and 06DR9059D (Germany), by GSI (TKrue 1012, GI/ME3,
OF/STR), by the Helmholtz Alliance HA216/EMMI, by the Excellence
Cluster Universe (Germany), by grants GAASCR IAA100480803 and MSMT
LC07050 (Czech Republic), by grant KBN5P03B
14020 (Poland), by INFN (Italy), by CNRS/IN2P3 (France),
by grants MCYT FPA2000-2041-C02-02 and XUGA PGID
FPA2009-12931 T02PXIC20605PN (Spain), by grant UCY-
10.3.11.12 (Cyprus), by INTAS grant 06-1000012-8861 and
EU contract RII3-CT-506078.
}




\begin{thebibliography}{99}
\bibitem{LeupoldMetagMosel}
S.~Leupold, V.~Metag, and U.~Mosel, Int. J. Mod. Phys. E \textbf{19}, 147 (2010).

\bibitem{GaleKapusta}
C.~Gale and J.~I.~Kapusta, Nucl. Phys. B \textbf{357}, 65 (1991).

\bibitem{CBM_physics_book}
B. Friman \etal\ (Eds), {\em The CBM Physics Book}, Lecture Notes in
Physics \textbf{814}, (Springer 2011).

\bibitem{ceres_pb_au_158_7per_centrality}
D.~Adamova \etal\ (CERES~Collab.), Phys. Lett. B \textbf{666}, 425 (2006).

\bibitem{na60_prl}
R.~Arnaldi \etal\ (NA60~Collab.), Phys. Rev. Lett. \textbf{96}, 162302 (2006).

\bibitem{dls_prl_porter}
R.~J.~Porter \etal\ (DLS~Collab.), Phys. Rev. Lett. \textbf{79}, 1229 (1997).

\bibitem{hades_tech}
G.~Agakishiev \etal\ (HADES~Collab.), Eur. Phys. J.~A \textbf{41}, 243 (2009).

\bibitem{hades_cc1gev}
G.~Agakishiev \etal\ (HADES~Collab.), Phys. Lett. B \textbf{663}, 43 (2008).

\bibitem{hades_cc2gev}
G.~Agakishiev \etal\ (HADES~Collab.), Phys. Rev. Lett. \textbf{98}, 052302 (2007).

\bibitem{taps_CC}
R.~Averbeck \etal\ (TAPS~Collab.), Z. Phys. A \textbf{359}, 65 (1997).

\bibitem{taps_CaCa}
R.~Holzmann \etal\ (TAPS~Collab.), Phys. Rev. C \textbf{56}, R2920 (1997).

\bibitem{taps_thermal}
R.~Averbeck, R.~Holzmann, V.~Metag, and R.~S.~Simon,
Phys. Rev. C \textbf{67}, 024903 (2003).

\bibitem{gale_kapusta}
C.~Gale and J.~I.~Kapusta, Phys. Rev. C \textbf{35}, 2107 (1987).

\bibitem{schaefer}
M.~Sch\"{a}fer, H.~C.~D\"{o}nges, A.~ Engel, and U.~Mosel,
Nucl. Phys. A \textbf{575}, 429 (1994).

\bibitem{shyam1}
R.~Shyam and U.~Mosel, Phys. Rev. C \textbf{67}, 065202 (2003).

\bibitem{kaptari}
L.~P.~Kaptari and B.~K\"{a}mpfer, Nucl. Phys. A \textbf{764}, 338 (2006);
Phys. Rev. C \textbf{80}, 064003 (2009).

\bibitem{dls_wilson}
W.~Wilson \etal\ (DLS~Collab.), Phys. Rev. C \textbf{57}, 1865 (1998).

\bibitem{hades_nn}
G.~Agakishiev \etal\ (HADES~Collab.), Phys. Lett. B \textbf{690}, 118 (2010).

\bibitem{shyam3}
R.~Shyam and U.~Mosel, arXiv:1006.3873v1.

\bibitem{hades_pi}
P.~Tlust\'{y} \etal\ (HADES Collab.), Bormio meeting 2009, arXiv:0906.2309v1 [nucle-ex].

\bibitem{hades_k0s}
G.~Agakishiev \etal\ (HADES~Collab.), Phys. Rev. C \textbf{82}, 044907 (2010).

\bibitem{simon_thesis}
S.~Lang, doctoral thesis, Frankfurt (2008).

\bibitem{tmva} Toolkit for MultiVariate Analysis with ROOT, http://tmva.sourceforge.net/.

\bibitem{martin_thesis}
M.~Jurkovic, doctoral thesis, M\"{u}nchen (2010).

\bibitem{filip_thesis}
F.~Krizek, doctoral thesis, Prague (2009).

\bibitem{pluto}
I.~Fr\"{o}hlich \etal, PoS (ACAT) 076 (2007).

\bibitem{pluto2}
F.~Dohrmann \etal, Eur. Phys. J. A \textbf{45}, 401 (2010).

\bibitem{glauber}
M.~L.~Miller, K.~Reygers, S.~J.~Sanders, and P.~Steinberg,
Annu. Rev. Nucl. Part. Sci. \textbf{57}, 205 (2007).

\bibitem{taps_AuAu}
A.~R.~Wolf \etal, Phys. Rev. Lett. \textbf{80}, 5281 (1998).

\bibitem{Reisdorf}
W.~Reisdorf \etal\ (FOPI Collab.), Nucl. Phys. A \textbf{781}, 59 (2007).

\bibitem{taps_omega1}
D.~Trnka \etal\ (CBELSA/TAPS~Collab.), Phys. Rev. Lett. \textbf{94}, 192303 (2005).

\bibitem{taps_omega2}
M.~Kotulla \etal\ (CBELSA/TAPS Collab.) Phys. Rev. Lett. \textbf{100}, 192302 (2008).

\bibitem{taps_omega3}
M.~Nanova \etal\ (CBELSA/TAPS Collab.) Phys. Rev. C \textbf{82}, 035209 (2010).

\bibitem{pdg2010}
K.~Nakamura \etal\ (Particle Data Group), J. Phys. G \textbf{37}, 075021 (2010).

\bibitem{hades_phi}
G.~Agakishiev \etal\ (HADES Collab.), Phys. Rev. C \textbf{80}, 025209 (2009).

\bibitem{hades_Xi}
G.~Agakishiev \etal\ (HADES Collab.), Phys. Rev. Lett. \textbf{103}, 132301 (2009).

\bibitem{hades_Lambda}
G.~Agakishiev \etal\ (HADES Collab.), Eur. Phys. J. A \textbf{47}, 21 (2011).

\bibitem{thermus}
S.~Wheaton and J.~Cleymans, Comput. Phys. Commun. \textbf{180}, 84 (2009).

\bibitem{OZI_review}
A.~I. Titov, B.~K\"{a}mpfer, and B.~L.~Reznik,
Phys. Rev. C \textbf{65}, 065202 (2002).

\bibitem{OZI_NN}
A.~Sibirtsev, Nucl. Phys. A \textbf{604}, 455 (1996).

\bibitem{OZI_piN}
A.~Sibirtsev and W.~Cassing, Eur. Phys. J. A \textbf{7}, 407 (2000).

\bibitem{disto}
F.~Balestra \etal\ (DISTO Collab.), Phys. Rev. C \textbf{63}, 024004 (2001).

\bibitem{anke1}
M.~Hartmann \etal\ (ANKE Collab.), Phys. Rev. Lett. \textbf{96}, 242301 (2006).


\bibitem{baldini}
A~Baldini, V.~Flaminio, W.~G.~Moorhead, and D.~R.~O.~Morrison,
in Landolt-B\"{o}rnstein, New Series \textbf{I/12a}, (Springer 1998).

\bibitem{cosytof}
M.~Abdel-Bary \etal\ (COSY-TOF Collab.), Phys. Lett. B \textbf{647}, 351 (2007).

\bibitem{BUU_phi}
H.~Schade, Gy.~Wolf, and B.~K\"{a}mpfer, Phys. Rev. C \textbf{81}, 034902 (2010).

\bibitem{catalytic_phi}
E.~E.~Kolomeitsev and B.~Tomasik, J. Phys. G \textbf{36}, 095104 (2009).

\bibitem{hagedorn}
R.~Hagedorn, Nuovo Cimento Suppl. \textbf{3}, 147 (1965). 

\bibitem{benjamin_thesis}
B.~Sailer, doctoral thesis, M\"{u}nchen (2007).

\bibitem{filip_qm09}
F.~Krizek \etal\ (HADES~Collab.), Nucl. Phys. A \textbf{830}, 483c (2009).

\bibitem{hades_CCpions}
G.~Agakishiev \etal\ (HADES~Collab.), Eur. Phys. J. A \textbf{40}, 45 (2009).

\bibitem{Reisdorf2}
W.~Reisdorf \etal\ (FOPI Collab.), Nucl. Phys. A \textbf{848}, 366 (2010).

\bibitem{na60_thermal}
R.~Arnaldi \etal\ (NA60 Collab.), Eur. Phys. J. C \textbf{61}, 711 (2009).

\bibitem{omega_abs}
G.~I.~Lykasov, W.~Cassing, S.~Sibirtsev, and M.~V.~Rzjanin,
Eur. Phys. J. A \textbf{6}, 71 (1999).

\bibitem{CLAS_omega}
M.~H.~Wood \etal\ (CLAS Collab.), Phys. Rev. Lett. \textbf{105}, 112301 (2010).

\bibitem{SPring8_phi}
T.~Ishikawa \etal\ (SPring8 Collab.), Phys. Lett. B \textbf{608}, 215 (2005).

\bibitem{faessler}
M~I.~Krivoruchenko, B~V.~Martemyanov, A.~Faessler, and C.~Fuchs,
Ann. Phys. \textbf{296}, 299 (2002).

\bibitem{iachello}
F.~Iachello, Workshop on $e^+e^-$ in the 1-2 GeV range, Alghero, Italy (2003), arXiv:nucl-th/0312074.


\bibitem{JacobWick}
M.~Jacob and G.~C.~Wick, Ann. Phys. \textbf{7}, 404 (1959).

\bibitem{LamTung}
C.~S.~Lam and W.~K.~Tung, Phys. Rev. D \textbf{18}, 2447 (1978).

\bibitem{kroll-wada}
N.~M.~Kroll and W.~Wada,  Phys. Rev. \textbf{98}, 1355 (1955).

\bibitem{landsberg}
L.~G.~Landsberg, Phys. Rep. \textbf{128}, 301 (1985).

\bibitem{samios}
N.~P.~Samios, Phys. Rev. \textbf{121}, 275 (1961).

\bibitem{hades_helicity1}
B.~Ramstein \etal\ (HADES~Collab.), Acta Physica Polonica B \textbf{41}, 365 (2009).

\bibitem{bratkovskaya_helicity1}
E.~L.~Bratkovskaya, O.~V.~Teryaev, and O.~V.~Toneev, Phys. Lett. B \textbf{348}, 283 (1995).

\bibitem{hades_helicity2}
B.~Ramstein \etal\ (HADES~Collab.), HADRON 2009: XIII International
Conference on Hadron Spectroscopy, Tallahassee (FL),
AIP Conf. Proc. \textbf{1257}, 695 (2010).

\bibitem{bratkovskaya_helicity2}
E.~L.~Bratkovskaya, M~Sch\"{a}fer, W.~Cassing, U.~Mosel, O.~V.~Teryaev,
and O.~V.~Toneev, Phys. Lett. B \textbf{348}, 325 (1995).

\bibitem{bratkovskaya_helicity3}
E.~L.~Bratkovskaya, W.~Cassing, and U.~Mosel, Phys. Lett. B \textbf{376}, 12 (1996);
Z. Phys. C \textbf{75}, 119 (1997).

\bibitem{gulamov_helicity}
T.~I.~Gulamov, A.~I.~Titov, and B.~K\"{a}mpfer, Phys. Lett. B \textbf{372}, 187
(1996).

\bibitem{na60_helicity}
R.~Arnaldi \etal\ (NA60 Collab.), Phys. Rev. Lett. \textbf{102}, 222301 (2009);
Eur. Phys. J. C \textbf{64}, 1 (2009).

\bibitem{martin_bormio10}
M.~Jurkovic \etal\ (HADES Collab.), XLVIII International Winter Meeting on
Nuclear Physics, Bormio 2010, PoS (BORMIO2010) 051.

\bibitem{bratkovskaya_arkcl}
E.~L.~Bratkovskaya and W.~Cassing, Nucl. Phys. A \textbf{807}, 214 (2008).

\bibitem{UrQMD09}
K.~Schmidt \etal, Phys. Rev. C \textbf{79}, 064908 (2009).

\bibitem{hades_upgrade}
J.~Pietraszko \etal,  in GSI Annual Report 2010.




\end{thebibliography}
\end{document}